\documentclass[fleqn,usenatbib]{mnras}
\PassOptionsToPackage{pdfpagelabels=false}{hyperref}
\pdfoutput=1
\usepackage{graphicx,color}
\usepackage{epstopdf}
\usepackage[latin1]{inputenc}
\usepackage[T1]{fontenc}
\usepackage{ae,aecompl}
\usepackage{graphics}
\usepackage{amsfonts}
\usepackage{amsmath}
\usepackage{multicol}
\usepackage{layout}
\usepackage{amssymb}
\usepackage[english]{babel}
\usepackage{scalerel}
\usepackage{times}
\usepackage{xcolor}
\newcommand\reallywidehat[1]{\arraycolsep=0pt\relax
\begin{array}{c}
\stretchto{
  \scaleto{
    \scalerel*[\widthof{\ensuremath{#1}}]{\kern-.5pt\bigwedge\kern-.5pt}
    {\rule[-\textheight/2]{1ex}{\textheight}} 
  }{\textheight} %
}{0.5ex}\\           
#1\\                 
\rule{-1ex}{0ex}
\end{array}
}
\title[Fast Weak Lensing 2.0]{
Testing the Reliability of Fast Methods for Weak Lensing Simulations:
{\sc wl-moka} on {\sc pinocchio}
}
\author[Giocoli C. et al. 2020]{\parbox{\textwidth}{
    Carlo Giocoli$^{1,2,3,4}$\thanks{E-mail:\href{mailto:carlo.giocoli@unibo.it}
      {carlo.giocoli@unibo.it}}, 
      Pierluigi Monaco$^{5,6,7,8}$,
      Lauro Moscardini$^{1,3,4}$
      Tiago Castro$^{5,6,7,8}$, 
      Massimo Meneghetti$^{3,4}$, 
      R. Benton Metcalf$^{1,2}$,      
      Marco Baldi$^{1,3,4}$
      }\\
      \\
$^{1}$Dipartimento di Fisica e Astronomia, Alma Mater Studiorum Universit\`{a} di Bologna, via Gobetti 93/2, I-40129 Bologna, Italy\\
$^{2}$Dipartimento di Fisica e Scienza della Terra, Universit\`a degli Studi di Ferrara, via Saragat 1, I-44122 Ferrara, Italy \\
$^{3}$INAF - Astrophysics and Space Science Observatory Bologna, via Gobetti 93/3, I-40129, Bologna, Italy \\
$^{4}$INFN - Sezione di Bologna, viale Berti Pichat 6/2, I-40127 Bologna, Italy \\
$^{5}$Dipartimento di Fisica, Sezione di Astronomia, Universit\`a di Trieste, Via Tiepolo 11, I-34143 Trieste, Italy\\
$^{6}$INAF -- Osservatorio Astronomico di Trieste, via Tiepolo 11, I-34131 Trieste, Italy\\
$^{7}$IFPU -- Institute for Fundamental Physics of the Universe, via Beirut 2, I-34151, Trieste, Italy\\
$^{8}$INFN -- Sezione di Trieste, I-34100 Trieste, Italy\\
}
\pubyear{2020}
\hypersetup{draft}
\begin{document}
\label{firstpage}
\pagerange{\pageref{firstpage}--\pageref{lastpage}}
\maketitle
\begin{abstract}

The generation of simulated convergence maps is of key importance in
fully exploiting weak lensing by Large Scale Structure (LSS) from
which cosmological parameters can be derived. In this paper we present
an extension of the {\sc pinocchio} code which produces catalogues of
dark matter haloes so that it is capable of simulating weak lensing by
LSS. Like {\sc wl-moka}, the method starts with a random realisation
of cosmological initial conditions, creates a halo catalogue and
projects it onto the past-light-cone, and paints in haloes assuming
parametric models for the mass density distribution within them.
Large scale modes that are not accounted for by the haloes are
constructed using linear theory.  We discuss the systematic errors
affecting the convergence power spectra when Lagrangian Perturbation
Theory at increasing order is used to displace the haloes within {\sc
  pinocchio}, and how they depend on the grid resolution.  Our
approximate method is shown to be very fast when compared to full
ray-tracing simulations from an N-Body run and able to recover the
weak lensing signal, at different redshifts, with a few percent
accuracy.  It also allows for quickly constructing weak lensing
covariance matrices, complementing {\sc pinocchio}'s ability of
generating the cluster mass function and galaxy clustering covariances
and thus paving the way for calculating cross covariances between the
different probes.  This work advances these approximate methods as
tools for simulating and analysing surveys data for cosmological
purposes.

\end{abstract}
\begin{keywords}
galaxies: haloes - cosmology: theory - dark matter - methods: analytic
- gravitational lending: weak
\end{keywords}

\section{Introduction}

Recent observational campaigns dedicated to the study of the
distribution of matter on large scales such as the ones coming from
the Cosmic Microwave Background (CMB) fluctuations
\citep{wmap9,planck1_14,planck16a}, cosmic shear
\citep{erben12,kilbinger13,hildebrandt17} and galaxy clustering
\citep{cole05,eisenstein05,sanchez14} tend to favour the so-called
standard cosmological model, where the energy-density of our Universe
is dominated by two unknown forms: Dark Matter and Dark Energy
\citep{peebles80,peebles93}.  This model successfully predicts
different aspects of structure formation processes
\citep{white78,baugh06,somerville15} going from the clustering of
galaxies on very large scales \citep{zehavi11,marulli13,beutler14} to
galaxy clusters
\citep{meneghetti08,postman12,meneghetti14,merten15,bergamini19}, to
the properties of dwarf galaxies
\citep{wilkinson06,madau08a,sawala15,wetzel16}.

Several experiments have been designed to constrain the cosmological
parameters with percent accuracy, however some of them have revealed
unexpected inconsistencies \citep{planckxxiv}.  In particular, while
the CMB temperature fluctuations probe the high redshift Universe,
gravitational lensing by large scale structures and cluster counts are
sensitive to low redshift density fluctuations. Recent comparisons
between high and low redshift probes have shown some differences in
the measured amplitude of the density fluctuations expressed through
the parameter $\sigma_8$: CMB power spectrum from Planck prefers a
slightly higher value of $\sigma_8$ with respect to the ones coming
from cosmic shear and cluster counts.

Gravitational lensing is a fundamental tool to study and map the
matter density distribution in our Universe
\citep{bartelmann01,kilbinger14}. For instance, while galaxy
clustering measurements probe the matter density field subject to the
galaxy bias \citep{sanchez12,marulli13,sanchez14,percival14,lee19},
tomographic lensing analyses opens the possibility of reconstructing
the projected total matter density distribution as a function of
redshift, and thus trace the cosmic structure formation in time
\citep{benjamin13,kitching14,hildebrandt17,kitching19}.  Since lensing
is sensitive to the total matter density present between the source
and the observer, it does not rely on any assumptions about the
correlation between luminous and dark matter. It is also very
sensitive to the presence of massive neutrinos as they tend to
suppress the growth of density fluctuations
\citep{lesgourgues06,massara14,castorina14,carbone16,poulin18}, thus
introducing a degeneracy with $\sigma_8$.

In order to better understand said inconsistency between low and high
redshift probes --- e.g. whether it comes from new physics or due to
systematics in the data analyses --- more dedicated measurements are
needed to reduce the statistical error-bars and possibly reveal a
significant tension.  For instance, weak gravitational lensing caused
by large scale structures, usually dubbed cosmic shear, will represent
one of the primary cosmological probes of various future wide field
surveys, like for example the ESA Euclid mission \citep{euclidredbook}
and LSST \citep{anderson01,ivezic09,lsst}.  When the number of
background sources, used to derive the lensing signal, is large, the
reconstruction of cosmological parameters depends mainly on the
control we have on systematics and covariances down to the typical
scale that is probed by the weak gravitational lensing
measurements. The construction of the covariance matrix requires the
production of a large sample of realisations that are able to not only
take into account all possible effects expected to be found in
observations but also to mimic as closely as possible the actual
survey.

In this work we present an extension of the latest version of {\sc
  pinocchio} \citep{munari17} which starts from the halo catalogue
constructed within a Past-Light-Cone (hereafter PLC) and simulates the
weak lensing signal generated by the intervening matter density
distribution up to a given source redshift. We have interfaced the PLC
output with {\sc wl-moka} \citep{giocoli17} in order to construct the
convergence map due to the intervening haloes. These algorithms
together reduce the computational cost of simulating the PLC on
cosmological scales by more than one order of magnitude with respect
to other methods based on N-body simulations, allowing the
construction of a very large sample of simulated weak lensing past
light cones to derive covariance matrices and, in a future work, also
to inspect the cosmological dependence of covariances.

The paper is organised as follows. In Sec. \ref{metandsec} we
introduce our method presenting the simulation data-set. In
Sec. \ref{results} we present our light-cone simulations and power
spectrum measurements. We summarise and conclude in
Sec. \ref{summary}.

\section{Methods and Simulations}
\label{metandsec}

In this section we describe the reference cosmological numerical
simulation with which we compare our approximate methods for weak
gravitational lensing simulations, and present our algorithms.

\begin{figure*}
  \includegraphics[width=0.51\hsize]{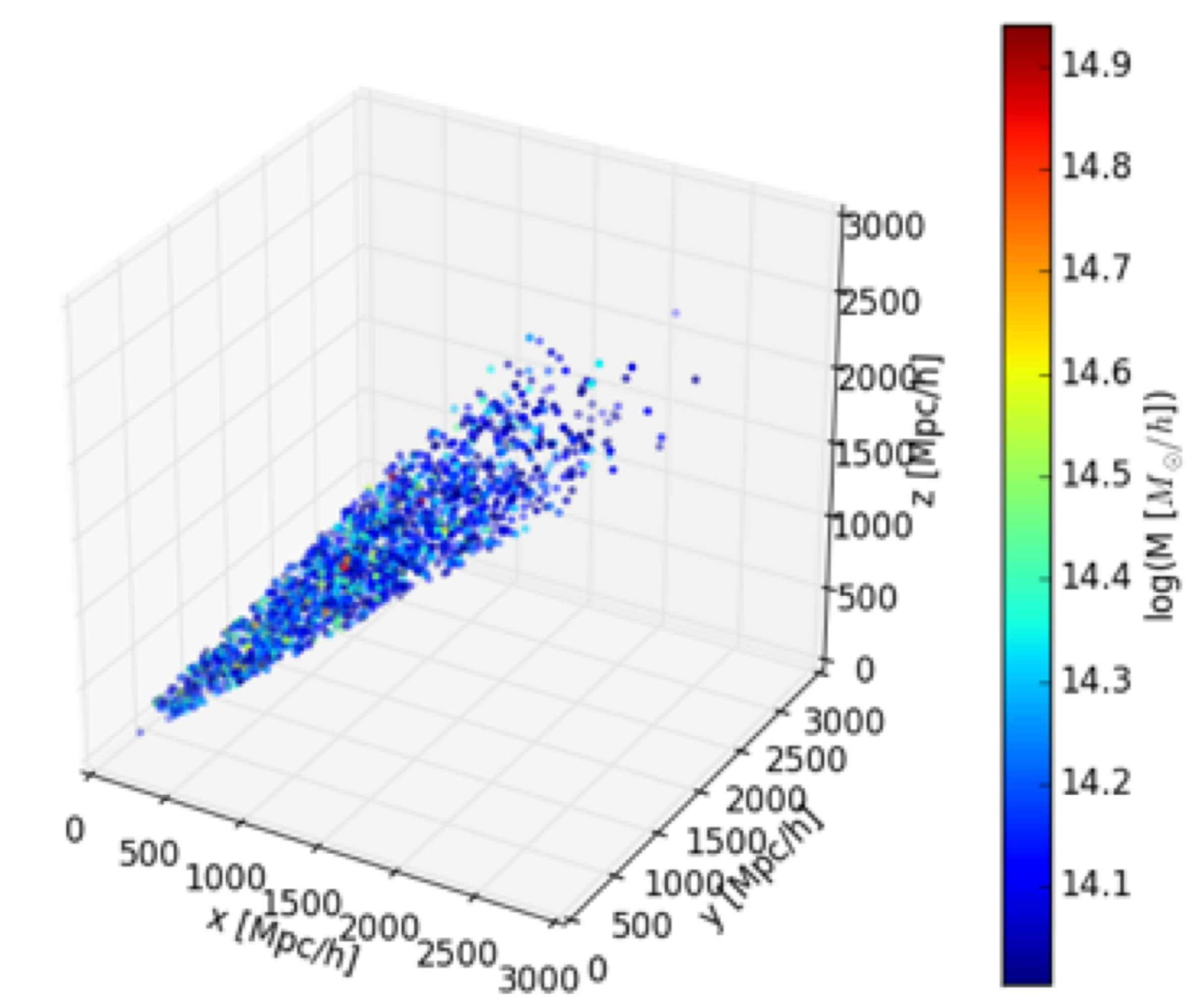}
  \includegraphics[width=0.45\hsize]{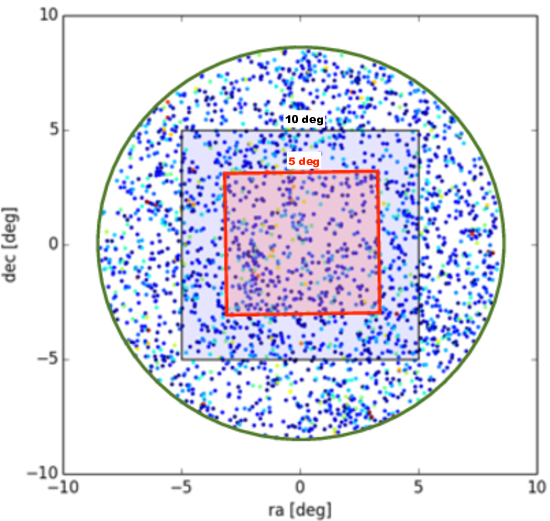}
  \caption{\label{figcone}Geometrical construction of the comoving PLC
    in {\sc pinocchio}.  In the left panel we show the comoving
    distribution of haloes with mass larger than $10^{14}M_{\odot}/h$
    up to redshift $z=4$ present within one realisation. The different
    masses are colour coded as indicated by the colour bar.  In the
    right panel we display the two-dimensional distribution of those
    haloes on the plane of the sky in angular coordinates. In this
    case the choice we have made about the angular geometry of the
    cone is more visible.  The semi-aperture has been set to $7.1$ deg
    which ensures that a square convergence map with $5$ deg on a side
    can be simulated.  This allows us to take into account the lensing
    contribution from haloes outside the field-of-view, from a buffer
    region, and reduce border effects.}
\end{figure*}

\subsection{Weak Lensing Maps from N-Body Simulations}

In this work we have used the $\mathrm{\Lambda}$CDM run of the CoDECS
project \citep[][]{baldi12b} as our reference N-Body simulation. The
run has been performed using a modified version of the widely used
TreePM/SPH N-body code GADGET \citep{springel05a} developed in
\citet{baldi10a}\footnote{In \citet{baldi10a} GADGET has been extended
  to cosmological scenarios with non-minimal couplings between Dark
  Energy and Cold Dark Matter particles. In this work we are limiting
  our analysis to the standard vanilla-$\mathrm{\Lambda}$CDM
  cosmology, thereby employing only the reference
  $\mathrm{\Lambda}$CDM run of the {\small CoDECS} simulations
  suite.}.  The {\small CoDECS} simulations adopt the following
cosmological parameters, consistent with the WMAP7 constraints by
\citet{komatsu11}: $\Omega_{\rm CDM} = 0.226$, $\Omega_b=0.0451$,
$\Omega_{\Lambda} = 0.729$, $h=0.703$ and $n_s=0.966$, with the
initial amplitude of linear scalar perturbations at CMB time ($z_{\rm
  CMB}\approx 1100$) set to $\mathcal{A}_s(z_{\rm CMB}) = 2.42 \times
10^{-9}$, resulting in a value of $\sigma_8=0.809$ at $z=0$.

The N-body run follows the evolution $2 \times 1024^3$ particles
evolved through collision-less dynamics from $z = 99$ to $z = 0$ in a
comoving box of $1$ Gpc/$h$ by side.  The mass resolution is $m_{\rm
  CDM} = 5.84 \times 10^{10} \mathrm{M_{\odot}}/h$ for the cold dark
matter component and $m_b = 1.17 \times 10^{10}\mathrm{M_{\odot}}/h$
for baryons, while the gravitational softening was set to $\epsilon_g
= 20\;\mathrm{kpc}/h$.  Despite the presence of baryonic particles
this simulation does not include hydrodynamics and is therefore a
purely collision-less N-body run. This is due to the original purpose
of the {\small CoDECS} simulations to capture the non-universal
coupling of a light dark energy scalar field to Dark Matter particles
only, leaving baryons uncoupled. Clearly, for the reference
$\mathrm{\Lambda}$CDM run -- where the coupling is set to zero -- no
difference is to be expected in the gravitational evolution of the
Dark Matter and baryonic components, which should be considered just
as two families of collision-less particles.

To build the lensing maps for light-cone simulations we stacked
together different slices of the simulation snapshots up to $z_s=4$.
We have constructed the PLC to have an angular squared aperture of $5$
deg on a side, which combined with the comoving size of the simulation
box of $1$ Gpc/$h$, ensures that the mass density distribution in the
cone has no gaps.  By construction, the geometry of the PLC is a
pyramid with a squared base, where the observer is located at the
vertex of the solid figure, while the final source redshift is placed
at the base.  In stacking up the various simulation snapshots and
collapsing them into projected particle lens planes we make use of the
\textsc{MapSim} code \citep{giocoli15,tessore15,castro17,hilbert19}.
The code initialises the memory and the grid size of the maps and
reads particle positions within the desired field of view (in this
case, $5$ deg on a side) from single snapshot files, which reduces the
memory consumption significantly.  The algorithm builds up the lens
planes from the present time up to the highest source redshift,
selected to be $z_s=4$.  The number of required lens planes is decided
ahead of time in order to avoid gaps in the constructed
light-cones. The lens planes are built by mapping the particle
positions to the nearest predetermined plane, maintaining angular
positions, and then pixelizing the surface density using the
Triangular Shaped Cloud (TSC) mass assignment scheme
\citep{hockney88}.  The grid pixels are chosen to have the same
angular size on all planes, equal to $2048\times 2048$, which allows
for a resolution of $8.8$ arcsec per pixel. The lens planes have been
constructed each time a piece of simulation is taken from the stored
particle snapshots; their number and recurrence depend on the number
of snapshots stored while running the simulation.  In particular, in
running our simulation we have stored $17$ snapshots from $z\sim4$ to
$z=0$.  In~\citet{castro17} it has been shown that a similar number of
snapshots is enough to reconstruct the PLC up to $z\sim5$ with lensing
statistics changing by less than $1\%$ if more snapshots are used.

The selection and the randomisation of each snapshot is done as in
\citet{roncarelli07} and discussed in more details in
\citet{giocoli15}.  If the light-cone arrives at the border of a
simulation box before it reaches the redshift limit where the next
snapshot will be used, the box is re-randomised and the light-cone
extended through it again.  Once the lens planes are created the
lensing calculation itself is done using the ray-tracing
\textsc{glamer} pipeline \citep{metcalf14,petkova14}.  In order to
have various statistical samples, we have created $25$ light-cone
realisations. They can be treated as independent since they do not
contain the same structures along the line-of-sight, considering the
size of the simulation box to be $1$ Gpc/$h$ and the field of view of
$5$ deg on a side. However, it is worth mentioning that all the
light-cones are constructed from the same N-body run and that they
share the same random realisation of the initial conditions of the
Universe. Even if their reconstructed lensing signal on small scales
depends on matter that occurs in the field-of-view, the large scale
modes are established by the seed in the initial conditions set up
when running the numerical simulation.

\subsection{Approximate Past-Light-Cones using Pinocchio}

We will compare the lensing simulations performed using the N-Body
run, with the ones constructed from the halo catalogues built up using
a fast and approximate algorithm: {\sc pinocchio} (namely its version
4.1.1).

{\sc pinocchio} is an approximate, semi-analytic public
code\footnote{\href{https://github.com/pigimonaco/Pinocchio}{https://github.com/pigimonaco/Pinocchio}},
based on excursion-set theory, ellipsoidal collapse and Lagrangian
Perturbation Theory (LPT), that is able to predict the formation of
dark matter haloes, given a cosmological linear density field
generated on a grid, without running a full N-body simulation.  It was
presented in \cite{monaco02}, then extended in \cite{monaco13} and
\cite{munari17}. The code first generates a density contrast field on
a grid, then it Gaussian-smooths the density using several smoothing
radii and computes, using ellipsoidal collapse, the collapse time at
each grid point (particle), storing the earliest value.  Later, it
fragments the collapsed medium with an algorithm that mimics the
hierarchical formation of structure.  Dark matter haloes are displaced
to their final position using LPT. The user can choose which
perturbation order to adopt: Zel'dovich Approximation, second-order
(2LPT) or third-order (3LPT).

Outputs are given both at fixed times and on the light cone
\citep{munari17}: for each halo, and for a list of periodic
replications needed to tile the comoving volume of the light cone, the
code computes the time at which the object crosses the light cone, and
outputs its properties (mass, position, velocity) at that time.

Using {\sc pinocchio} we have produced different simulations as
summarised in Table \ref{tabsims}.  We set cosmology and box
properties identical to those used for the reference N-body
simulation, but with different initial seed numbers so as to have
several realisations of the same volume. This allows us to beat down
sample variance on the predicted convergence power spectrum. Our
reference \textsc{pinocchio} simulation has been run with $1024^3$
grid points and in the 3LPT configuration for the particle
displacements starting from the initial conditions. This run consists
of $512$ different realisations and corresponding past light-cones. We
have chosen the semi-aperture of the past light-cone to be $7.1$ deg
which gives a total area in the plane of the sky of $158.37$ deg$^2$.
This value guarantees us the possibility of creating a pyramidal
configuration for the convergence maps -- consistent with the maps
constructed from the N-body simulation -- with $5$ deg by side up to a
final source redshift $z_s=4$.  In addition to the reference
\textsc{pinocchio}$_{\rm 3LPT}$ we produced a sample of other
approximate simulations: $25$ using the same grid resolutions but
adopting both 2LPT and ZA displacements, $512$ with a lower resolution
grid ($512^3$) and 3LPT, and $25$ with a higher resolution grid
($2048^3$) and 3LPT displacements. All corresponding mass resolutions
are reported in the third column of Table \ref{tabsims}.  The random
numbers of the initial conditions for the various {\sc pinocchio}
simulations have been consistently chosen to be identical, in the
sense that the 512 low resolutions runs with $512^3$ grid size have
the same initial random displacement fields of the $1024^3$, and so
the $25$ runs performed with the different displacement fields or
using a higher resolution grid of $2048^3$ that share the initial
condition seeds with the first $25$ reference runs.  This will allow
us a more direct comparison between the different runs and convergence
maps starting from the same initial displacement field of the
theoretical linear power spectrum.

The typical CPU for a $1024^3$ in 1 Gpc/$h$ N-Body simulations from
$z=99$ to $z=0$, plus i/o $90$ snapshots and on-the-fly halo finding
procedures is approximately $50.000$ CPU hours; plus $1.000$ more
hours for light-cone productions and multi-plane ray-tracing, a
typical super architecture like the ones available at
CINECA\footnote{http://www.hpc.cineca.it/content/hardware}.  As for
the required resources, a \textsc{pinocchio} run with $1024^3$
particles demands a CPU time of order of $10$ hours on a
supercomputer; the version with Zel'dovich or 2LPT displacements can
fit into a single node with 256 Gb of RAM, while the 3LPT version will
require more memory; the elapsed time will be of order of 10 min in
this case. Higher orders require a few more FFTs, for an overhead of
order of $\sim$10\% going from Zel'dovich to 3LPT.  The light-cone
on-the-fly construction requires an even smaller overhead, $\sim$4\%
for the configuration used in this paper.  The scaling was
demonstrated in \cite{munari17} to be very similar to $N\log_2 N$, so
going from $512^3$ to $1024^3$, or from this to $2048^3$, requires a
factor of $\sim9$ more computing time and a factor of 8 more memory;
if the number of used cores increases as the RAM, the wall-clock time
will not change much. Painting haloes on the PLC halo catalogues takes
not more than $2$ CPU hours per light-cone, scaling with the number
density of systems that depends on the minimum mass threshold
considered and on maximum source redshift. This means a ratio of CPU
times between full N-Body and Fast Approximate methods in producing
one convergence map of $5$ deg by side of approximately $3\times
10^3$. To these, we have to highlight the fact that each {\sc
  pinocchio} light-cone has the advantage of having a different
Initial Condition (IC) set up, while this is not the case for
past-light-cones extracted from the N-body.

\begin{table*}
  \caption{Summary of the simulations.  The symbol * marks our
    reference \textsc{pinocchio} run.  For the N-Body case it is worth
    mentioning that all the various light-cones have been generated
    from the same cosmological simulation, randomising the various
    snapshots using the \textsc{MapSim} code.  By construction the
    particle mass resolution of our reference \textsc{pinocchio} run
    is equal to that of the N-Body simulation; all the runs consider a
    cosmological box of 1 Gpc/$h$ comoving by side.\label{tabsims}}
  \begin{tabular}{l|c|c|c|c}
    $ $ & field-of-view [$\mathrm{deg^2}$] (haloes) & min. halo mass [$\mathrm{M_{\odot}/}h$] & n. real.  & field-of-view [$\mathrm{deg^2}$] (convergence)  \\ \hline
    (*) \textsc{pinocchio}$_{\mathrm{3LPC}}$ ($1024^3$) & 158.37 & $7.0 \times 10^{11}$ & 512 & 25  \\ \hline
    \textsc{pinocchio}$_{\mathrm{2LPC}}$ ($1024^3$) & 158.37 & $7.0 \times 10^{11}$ & 25 & 25  \\ \hline
    \textsc{pinocchio}$_{\mathrm{ZA}}$ ($1024^3$) & 158.37 & $7.0 \times 10^{11}$ & 25 & 25  \\ \hline
    \textsc{pinocchio}$_{\mathrm{3LPC}}$ ($512^3$) & 158.37 & $5.6 \times 10^{12}$ & 512 & 25  \\ \hline
    \textsc{pinocchio}$_{\mathrm{3LPC}}$ ($2048^3$) & 158.37 & $8.75 \times 10^{10}$ & 25 & 25  \\ \hline
    N-Body & 100 & $7.0 \times 10^{10}$ (total part. mass: DM + bar.) & 25 & 25  \\ \hline
  \end{tabular}
\end{table*}

In the left panel of Fig.~\ref{figcone} we show the comoving halo
distribution in the past-light-cones constructed by
\textsc{pinocchio}; haloes have different colour according to their
mass, as indicated by the colour-bar.  In the right panel we display
the two-dimensional distribution of haloes, in angular coordinates as
they appear in the plane of the sky up to $z=4$. In the right panel we
draw also the size of the squared postage-stamp of $5$ deg by side
representing the geometry of our final convergence map. The geometry
of \textsc{pinocchio} PLC allows us to consider the lensing
contribution from haloes outside the field-of-view, from a buffer
region, and accounting also for border effects: lensing signal due to
haloes which are not in the final field of $5$ deg by side.

\begin{figure}
  \includegraphics[width=\hsize]{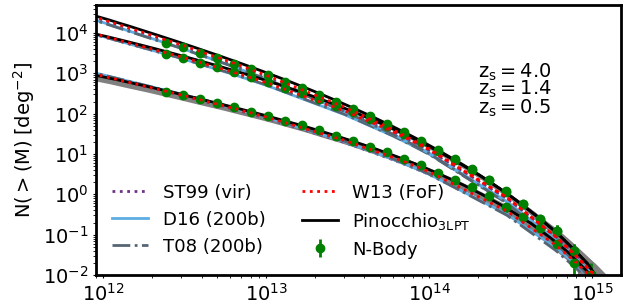}
  \includegraphics[width=\hsize]{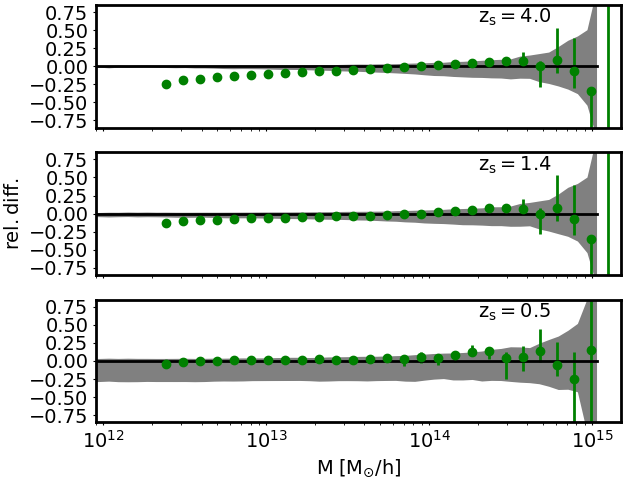}
  \caption{Cumulative halo mass function in unit of square degree
    within the constructed past-light-cones.  The solid black curves
    display the median over the various realisations, the shaded grey
    area encloses the quartiles.  The green data points indicate the
    median measurements from the N-Body simulation over $25$ different
    light-cones. The dotted magenta, solid light-blue, dot-dashed dark
    grey and red dotted display the predictions from \citet{sheth99b},
    \citet{despali16}, \citet{tinker08} and \citet{watson13} mass
    functions, respectively. The three bottom sub-panels show the
    relative difference between the mass function from the N-Body and
    \textsc{pinocchio} simulations, up to the three considered source
    redshifts.\label{figmf}}
\end{figure}

In Figure \ref{figmf} we show the cumulative halo mass function
normalised to a one square degree light-cone, from $z=0$ up to
redshift $0.5$, $1.4$ and $4$ from bottom to top, respectively. In
both cases haloes have been identified using a Friends-of-Friends
algorithm. It is worth to underline that in {\sc pinocchio} the
expression for the threshold distance that determines accretion and
merging includes free parameters \citep{munari17} as the linking
length parameter for the FoF definition in the N-Body simulations.
The comparison of the FoF mass functions between {\sc pinocchio} and
N-Body simulation is fair and consistent, indeed in both cases we use
the same methodology to find collapsed structures. For comparison the
dotted red curves display the FoF mass function as calibrate by
\citet{watson13}.  The lower panels display the relative differences
of the median counts computed averaging $25$ different simulation
light-cones of the N-Body run with respect to the average predictions
from the $512$ \textsc{pinocchio} simulations; the shaded area gives
the sample variance measured with the quartiles from
\textsc{pinocchio} runs, that is large at low $z$ due to the small
sampled volume. Only for comparison purpose, in the top panel we show
also the expectation from \citet{sheth99b}, where they use the virial
definition for the halo mass, and from \citet{tinker08,despali16}
assuming a threshold corresponding to $200$ times the comoving
background density, which has been shown to be the closest to the FoF
mass functions \citep{knebe11}.  In the lower panel we notice that
there is a very good agreement between the halo counts in the N-Body
and \textsc{pinocchio} light-cones down to $5\times
10^{12}M_{\odot}/h$, however the higher the redshift is the more the
N-body counts suffer from a small reduction toward small masses due to
particle and force resolutions. The error bars on the green data
points and the grey shaded regions enclosing the black lines bracket
the first and the third quartiles of the distribution at a fixed halo
mass.

\subsubsection{Weak lensing simulations using projected halo model}

In this section we introduce the lensing notations we will adopt
throughout the paper; the symbols and the equations are quite general
and consistent between the two methods adopted in constructing the
convergence maps from particles and haloes.

Defining ${\pmb \theta}$ the angular position on the sky and ${\pmb
  \beta}$ the position on the source plane (the unlensed position),
then a distortion matrix ${\bf A}$, in the weak lensing regime, can be
read as
\begin{align}
{\bf A} \equiv \frac{\partial {\pmb \beta}}{\partial {\pmb \theta} } =
\left(
\begin{array}{cc}
1-\kappa-\gamma_1 & \gamma_2  \\
\gamma_2  & 1-\kappa + \gamma_1
\end{array}
\right)\,,
\end{align}
where scalar $\kappa$ represents the convergence and the pseudo-vector
${\pmb \gamma}\equiv\gamma_1+i \gamma_2$ the shear tensor\footnote{In
  tensor notation we can read the shear as:\begin{align} \left(
\begin{array}{cc}
\gamma_1 & \gamma_2  \\
\gamma_2  & -\gamma_1
\end{array}
\right).
\end{align}}.  In the case of a
single lens plane, the convergence can be written as:
\begin{equation}
\kappa({\pmb \theta}) \equiv \frac{\Sigma({\pmb \theta})}{\Sigma_{\rm
    crit}}\, , \label{eqconvergence}
\end{equation}
where $\Sigma({\pmb \theta})$ represents the surface mass density and
$\Sigma_{\rm crit}$ the critical surface density:
\begin{equation}
\Sigma_{\rm crit} \equiv \dfrac{c^2}{4 \pi G} \dfrac{D_l}{D_s D_{ls}},
\end{equation}
where $c$ indicates the speed of light, $G$ the Newton's constant and
$D_l$, $D_s$ and $D_{ls}$ the angular diameter distances between
observer-lens, observer-source and source-lens, respectively.

\begin{figure*}
  \includegraphics[width=\hsize]{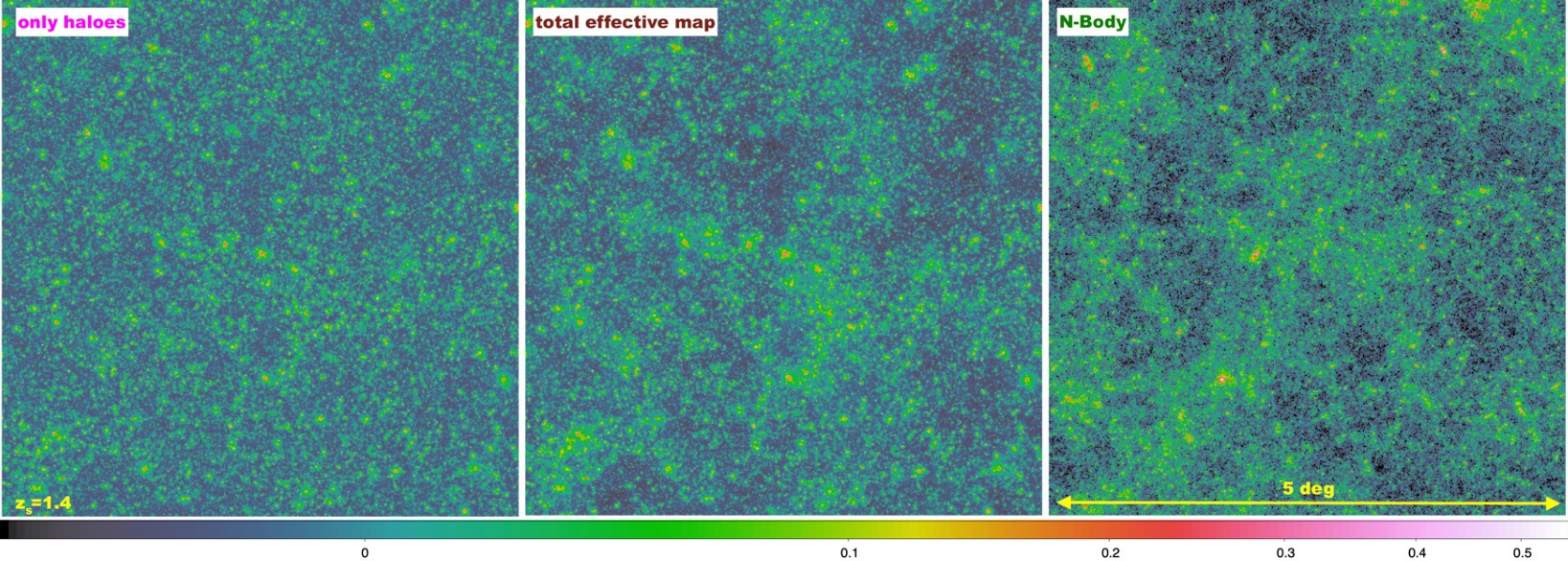}
  \caption{Convergence maps for sources at $z_s=1.4$ for a
    field-of-view of $5\times5$ square degrees. Left panel shows the
    map obtained from the halo catalogue of a \textsc{pinocchio}
    past-light-cone.  Central panel displays the effective total
    contribution of haloes plus not resolved in haloes, modelled using
    linear theory. Right panel shows, for comparison, the convergence
    map of a light-cone constructed using \textsc{MapSim} from the
    snapshots of a cosmological numerical simulation with a different
    realisation of the initial conditions.\label{figmap1}}
\end{figure*}

Following a general consensus, we will assume that matter in haloes is
distributed following the \citet{navarro96} (hereafter NFW) relation:
\begin{equation}
  \rho(r|M_{\rm h}) = \frac{\rho_s}{(r/r_s)(1+r/r_s)^2}\,,
\end{equation}
where $r_s$ is the scale radius, defining the concentration $c_{\rm h}
\equiv R_{\rm h}/r_s$ and $\rho_s$ the dark matter density at the
scale radius:
\begin{equation}
  \rho_s = \frac{M_{\rm h}}{4 \pi r_s^3}
  \left[ \ln(1+c_{\rm h}) - \frac{c_{\rm h}}{1+c_{\rm h}}\right]^{-1}\,,
\label{eqrhos}
\end{equation}
$R_{\rm h}$ is the radius of the halo which may vary depending on the
halo over-density definition.

From the hierarchical clustering model the halo concentration $c_{\rm
  h}$ is expected to be a decreasing function of the host halo
mass. Small haloes form first \citep{vandenbosch02,deboni16} when the
universe was denser and then merge together forming the more massive
ones: galaxy clusters sit at the peak of the hierarchical pyramid
being the most recent structures to form
\citep{bond91,lacey93,sheth04a,giocoli07}.  This trend is reflected in
the mass-concentration relation: at a given redshift smaller haloes
are more concentrated than larger ones.  Different fitting functions
for mass-concentration relations have been presented by various
authors
\citep{bullock01a,neto07,duffy08,gao08,meneghetti14,ragagnin19}.  In
this work, we adopt the relation proposed by \citet{zhao09} which
links the concentration of a given halo with the time $t_{0.04}$ at
which its main progenitor assembles $4$ percent of its mass.  For the
mass accretion history we adopt the model proposed by
\citet{giocoli12b} which allows us to trace back the full halo growth
history with cosmic time down to the desired time $t_{0.04}$. We want
to underline that the model by \citet{zhao09} also fits numerical
simulations with different cosmologies; it seems to be of reasonably
general validity within a few percent accuracy, as is the generalised
model of the mass accretion history we adopt as tested by
\citet{giocoli13}. It is interesting to notice that the particular
model for the concentration mass relation mainly impacts on the
behaviour of the power spectrum at scales below $1\,h^{-1}$Mpc as
discussed in details by \citet{giocoli10b}.  Due to different assembly
histories, haloes with the same mass at the same redshift may have
different concentrations
\citep{navarro96,jing00,wechsler02,zhao03a,zhao03b}.  At fixed halo
mass, the distribution in concentration is well described by a
log-normal distribution function with a rms $\sigma_{\ln c}$ between
$0.1$ and $0.25$ \citep{jing00,dolag04,sheth04b,neto07}. In this work
we adopt a log-normal distribution with $\sigma_{\ln c}=0.25$.  We
decided to follow this approach in assigning the halo concentration to
be as general as possible. Results from the analyses of various
numerical simulations have revealed that, at fixed halo mass,
structural properties, like concentration and subhalo population,
depend on the halo assembly histories
\citep{giocoli08b,giocoli10a,giocoli12b,lange19,zehavi19,montero-dorta20,yangyao20}.
However saving all those data three files of the halo catalogues would
have increased much the storage capability we planned for this
project. As test case in \citep{giocoli17}, we have also generated the
halo convergence maps reading the halo concentration from the
corresponding simulated N-Body catalogue finding that the assembly
bias effect on the convergence power spectra has only sub percent
effects.\\ As presented by \citep{bartelmann96a}, assuming spherical
symmetry the NFW profile has a well defined solution when integrated
along the line of sight up to the virial radius:
\begin{equation}
    \kappa(x_1,x_2|M_h,z_l,z_s) = 2 \int_0^{R_{\rm vir}} \rho(x_1,x_2,\zeta|M_h) \mathrm{d} \zeta / \Sigma_{crit}(z_l,z_s) \label{eqsigma}
\end{equation}
with $r^2 = x_1^2+x_2^2+\zeta^2$ \citep{giocoli12a,giocoli17}.
Expressing $\xi^2 = x_1^2 + x_2^2$ we can write:
\begin{equation}
\kappa_0 = 2.0/\Sigma_{crit}(z_l,z_s),
\end{equation}
and
\begin{equation}
\kappa(\xi) = \kappa_0 (\kappa_1 + \kappa_2 + \kappa_3)/ \kappa_D,
\end{equation}
with
\begin{equation}
\begin{cases}
\mathrm{for\;} \xi < 1:  \\
\kappa_1 = R_{vir} \sqrt{1 -\xi^2}\left(1 - \sqrt{\xi^2 + R_{vir}^2} \right) \\
\kappa_2 = - \dfrac{ \xi^2 + R_{vir}^2 - 1}{2} \log \left(\dfrac{\dfrac{R_{vir}}{\sqrt{1 - \xi^2}} + 1}{\dfrac{R_{vir}}{\sqrt{1 - \xi^2}} -1} \right)  \\
\kappa_3 = \dfrac{\xi^2 + R_{vir}^2 - 1}{2}\log \left(\dfrac{\left(\dfrac{R_{vir}}{\sqrt{1-\xi^2}\sqrt{\xi^2 + R_{vir}^2}}\right)+1}{\left( \dfrac{R_{vir}}{\sqrt{1-\xi^2}\sqrt{\xi^2 + R_{vir}^2}} \right) -1} \right)  \\
\kappa_D = \left(1-\xi^2\right)^{3/2}\left(\xi^2+R_{vir}^2-1\right) \\
\end{cases}
\end{equation}
\begin{equation}
\begin{cases}
\mathrm{for\;} \xi > 1:  \\
\kappa_1 = R_{vir} \sqrt{\xi^2-1}\left(\sqrt{\xi^2 + R_{vir}^2}-1 \right) \\
\kappa_2 = -\left( \xi^2 + R_{vir}^2 - 1 \right) \arctan \left(\dfrac{R_{vir}}{\sqrt{\xi^2-1}} \right)  \\
\kappa_3 = \left(\xi^2 + R_{vir}^2 - 1 \right) \arctan \left(\dfrac{R_{vir}}{\sqrt{\xi^2-1}\sqrt{\xi^2 + R_{vir}^2}}\right)  \\
\kappa_D = \left(\xi^2-1\right)^{3/2}\left(\xi^2+R_{vir}^2-1\right) \\
\end{cases}
\end{equation}
and
\begin{equation}
\begin{cases}
\mathrm{for\;} \xi = 1:  \\
\kappa(\xi) = \dfrac{\kappa_0}{3} \sqrt{1 + \dfrac{1}{R_{vir}^2}}.
\end{cases}
\end{equation}
The contribution to the convergence from each halo within the
field-of-view is modulated by the critical density that depends on the
observer-lens-source configuration, as we have expressed in the
equations above.

In the left panel of Fig.  \ref{figmap1} we show the convergence map
reconstructed using halo positions, masses and redshift from one PLC
realisation and assuming a fixed source redshift of $z_s=1.4$. We can
see the contribution from all the mass in the haloes and the presence
of galaxy clusters at the intersections of filaments. As discussed by
\citet{giocoli17} the reconstructed power spectrum using only haloes
fails in reproducing the expectation on large scales from linear
theory. This inconsistency is a manifestation of the absence of large
scale modes sampled by small mass haloes (below our mass resolution
threshold) and by the diffuse matter that is not in haloes. A
straightforward way to add this power back is to project on the past
light-cone particle positions that are outside haloes, construct
density planes in redshift bins and add them to those obtained with
haloes. This procedure is feasible and will be presented in a future
paper; however it implies significant overhead in CPU time and
storage: it requires writing particle properties to the disk,
something that is avoided by \textsc{pinocchio} in its standard
implementation. In the context of the massive generation of mock halo
catalogues, it is very convenient to adopt the procedure proposed by
\cite{giocoli17} to reconstruct the missing power from the halo
catalogue itself.  We have also estimated the effect of including in
the models the missing diffuse matter present between halos by
extending the truncation of the density profile at different values of
the virial radius. This effect mainly manifests in the transition
between the 1- and 2-halo term up to large scales and has not much
effect at the 1-halo level, in the projected power spectrum. The
extension of the halo density profiles outside the virial radii
creates a trend with the source redshifts, artificially increasing the
mean background density of the universe, due to the projected
intervening matter density distribution along the
line-of-sight. Nonetheless, we decided to be conservative in our
method, as typically done in the halo model formalism
\citep{cooray02}, assuming the collapsed matter in haloes only up to
the virial radius.

In order to include the large scale modes due to unresolved matter not
in haloes, we generate in Fourier space a field with a random Gaussian
realisation whose amplitude is modulated by $P_{\kappa, \rm lin}(l)$.
The phases are chosen to be coherent with the halo location within the
considered map \citep{giocoli17}.  By construction, summing the
convergence maps of the haloes with the one from the linear theory
contribution gives a two-dimensional map that includes the cross-talk
term between the two fields:
 \begin{eqnarray}
 \langle \hat{\kappa}(\mathbf{l}) \hat{k^*}(\mathbf{l}') \rangle &=&
  \langle \reallywidehat{(\kappa_{\rm hm} + \kappa_{\rm lin})}(\mathbf{l})
  \reallywidehat{(\kappa_{\rm hm} + \kappa_{\rm lin})^*}(\mathbf{l}) \rangle \\
  &=&  4 \pi^2 \delta_D(\mathbf{l}-\mathbf{l}') \left(
  P_{\kappa_{\rm hm}}(l) + P_{\kappa_{\rm lin}}(l) + P_{\rm hm-lin}(l)
  \right), \nonumber
 \end{eqnarray}
 where $P_{\rm hm-lin}(l)$ indicates the cross-spectrum term between
 the two fields and, by definition, $P_{\kappa_{\rm
     lin}}(l)=P_{\kappa_{\rm lin,r}}(l)$, where $P_{\kappa_{\rm
     lin,r}}(l)$ indicates the power spectrum of a map with random
 phases.  Because of the cross-spectrum term, we then re-normalise the
 map to match the large scale behaviour predicted on large scale by
 linear theory using the relation:
\begin{equation}
  A(l) = \dfrac{P_{\kappa,\rm lin}(l)}{P_{\kappa_{\rm hm}+\kappa_{\rm
        lin}}(l)}, \label{eqampl}
\end{equation}
where $P_{\kappa_{\rm hm}+\kappa_{\rm lin}}(l) = P_{\kappa_{\rm
    hm}}(l) + P_{\kappa_{\rm lin}}(l) + P_{\rm hm-lin}(l)$\footnote{We
  have also tested the case of subtracting the cross talk spectra to
  the total power. However, it is worth mentioning that this gives the
  same results.}.  Dividing the contributions to the convergence power
spectrum in haloes and diffuse component allows us to discriminate two
separate contributions.  While the diffuse matter, relevant on large
scales and treated using linear theory, represents the Gaussian
contribution to the power spectrum, the haloes, important on small
scales in the non linear regime, portrays the non-Gaussian stochastic
part. Haloes are non-linear regions of the matter density fluctuation
field disjoint from the expansion of the universe, their structural
properties -- concentration, substructures, density profiles etc --
shape the small scale modes
\citep{cooray02,smith03,sheth03c,giocoli10b}.

In the central panel of Fig.  \ref{figmap1} we display the convergence
map where we include also the modelling of the large scale modes
coherent with the halo distribution within the field of view.  In
order to do so we use as reference the prediction, in the Fourier
space, from the linear theory of the convergence power spectrum
$P_{\kappa, \rm lin}(l)$ that in the Born Approximation and for source
redshift at $z_s$ can be read as:
\begin{equation}
P_{\kappa, \rm lin}(l) = \dfrac{9 H_0^4 \Omega_m^2}{4 c^4} \int_0^{w_s(z_s)}
\left(\dfrac{D(z,z_s)}{D(z_s) a}\right)^2 P_{\rm \delta,lin}\left(l\dfrac{a}{D(z)},z\right) \, \mathrm{d}w,
\label{eqborn}
\end{equation}
where $H_0$ and $\Omega_m$ represent the present day Hubble constant
and matter density parameter, $c$ indicates the speed of light, $w(z)$
and $D(z)$ the radial comoving and angular diameter distances at
redshift $z$, $a \equiv 1/(1+z)$, and $P_{\rm \delta, lin}(k,z)$ the
linear matter power spectrum at a given comoving mode $k$ re-scaled by
the growth factor at redshift $z$.  For comparison, the third panel of
Fig. \ref{figmap1} displays the convergence map of the past-light-cone
constructed from the cosmological numerical simulation up to
$z_s=1.4$; we recall that the simulation {\em does not} trace the same
large-scale structure as the {\sc pinocchio} realisation.

\subsection{Halo Model for non-linear power spectrum}
\label{hmsection}

The non-linear matter density distribution for $k>1 h/$Mpc can be
reconstructed using the halo model formalism. This is based on the
assumption that all matter in the universe can be associated to
collapsed and virialised haloes. In real space the matter-matter
correlation can be decomposed into two components:
\begin{equation}
\xi(r) = \xi_{1h}(r) + \xi_{2h}(r)
\end{equation}
where $\xi_{1h}(r)$ and $\xi_{2h}(r)$ are term one and two halo
components, that account for the matter-matter correlation in the same
or in distant haloes, respectively. Following the halo model formalism
as described by \citep{cooray02,giocoli10b}, we can relate real and
Fourier, making explicit the redshift dependence, considering that:
\begin{equation}
 P(k,z) = 4 \pi \int \xi(r,z)\frac{\sin(kr)}{kr}r^{2} {\rm d}r \,,
\end{equation}
and write the one and two halo term in Fourier space as:
\begin{eqnarray}
 P_{1H}(k,z)  &=&   \int_{M_{\rm min}}  \left(\dfrac{M_h}{\bar{\rho}}\right)^2  n(M_h,z)
 \nonumber  \\   &\times&  \int  p(c_h|M_h)   u^2(k|c_h(M_h))  \mathrm{d}c_h  \,
 \mathrm{d}M_h\,, \label{pkhalo1}
\end{eqnarray}
\begin{eqnarray}
 P_{2H}(k,z)  &=& P_{\rm  \delta,lin}(k)  \Big{[} \int_{M_{\rm min}}  \dfrac{M_h}{\bar{\rho}}
   n(M_h,z)\,b(M_h,z)  \nonumber \\    &\times&          \int
   p(c_h|M_h)\,u(k|c_h,M_h)\,\mathrm{d}c_h
   \,\mathrm{d}M_h \Big{]}^2 \,  \label{pkhalo2}
\end{eqnarray}
where $u(k|c_h,M_h)$ represents the Fourier transform of the NFW
matter density profile, $b(M_h,z)$ the halo bias for which we use the
model by \citet{sheth99b} and $n(M_h,z)$ the halo mass function for
which we adopt the \citet{despali16} model that describes very well
the mass function of {\sc pinocchio} light-cones, as can be noticed in
Fig.~\ref{figmf}. In the two equations above we have made it explicit
that the halo mass function is typically integrated from a given
minimum halo mass $M_{\rm min}$ and that we use a stochastic model for
the concentration mass relation with a scatter $\sigma_{\ln c}=0.25$,
as considered in {\sc wl-moka}. It is worth mentioning that
eq.~\ref{pkhalo2} needs to be normalised by
\begin{equation}
 P_{2H,0}(k,z)  =  \Big{[} \int_{M_{\rm min}}  \dfrac{M_h}{\bar{\rho}}
   n(M_h,z)\,b(M_h,z) \,\mathrm{d}M_h \Big{]}^2 \,  \label{pkhalo20}
\end{equation}
since it has to match the linear theory on large scales, i.e. for
small $k$.

\begin{figure}
  \includegraphics[width=\hsize]{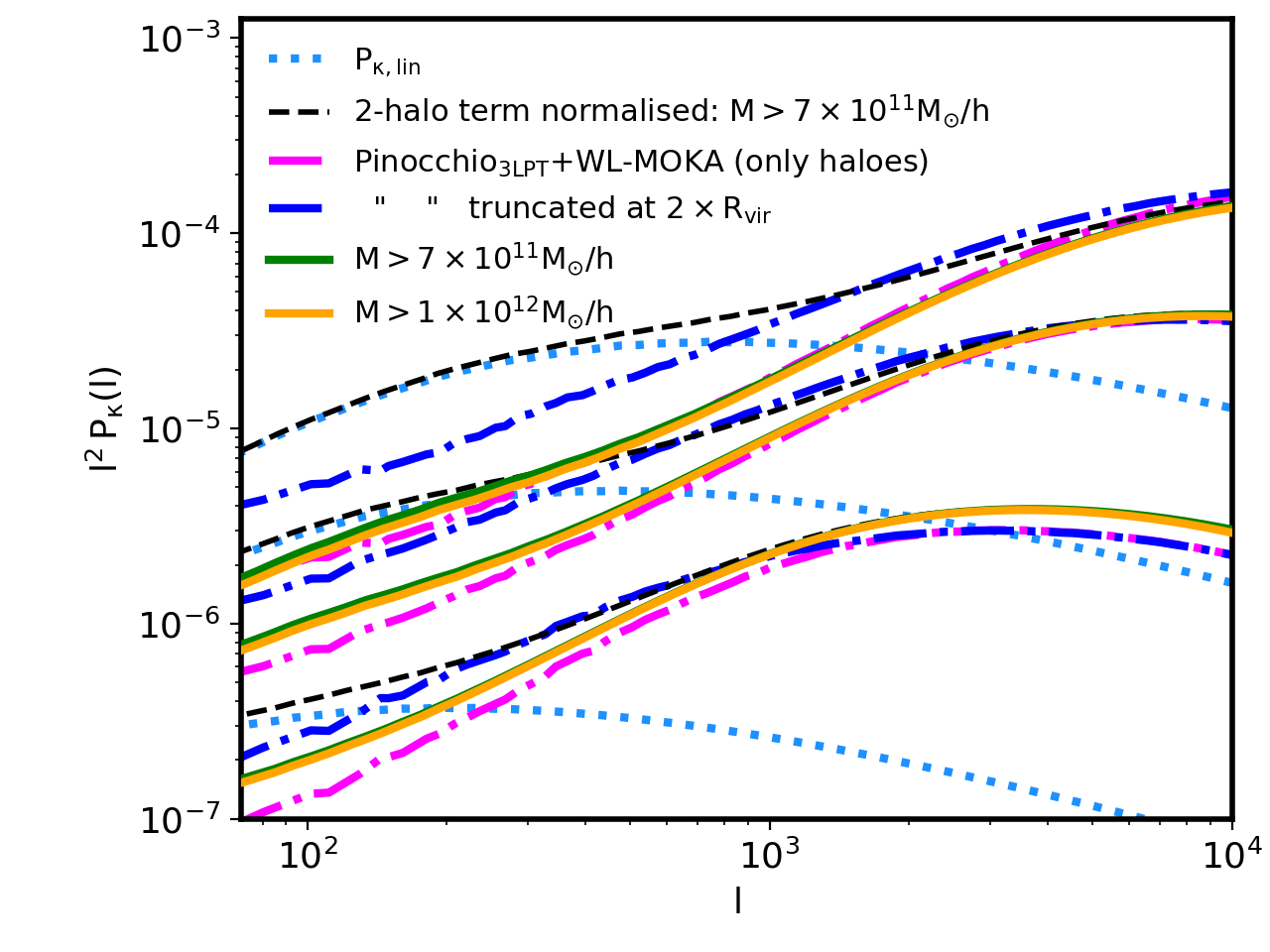}
  \caption{Convergence power spectrum prediction using halo model
    formalism. The black dashed curves show the prediction at three
    different fixed source redshifts using eq. (\ref{eqborn}) (from
    bottom to top, $z=0.5$, $1.4$ and $4$) where the lower limit of
    halo mass function integrals is $M_{\rm min} = 7 \times 10^{11}
    M_{\odot}/h$ and the two halo term has been normalised using
    eq.~(\ref{pkhalo20}) to match the linear prediction -- dotted
    light-blue curves. The magenta (blue) dot-dashed curves show the
    average measurements done on $512$ convergence maps constructed
    combining {\sc pinocchio} PLC and {\sc wl-moka} using only haloes
    (truncating the profile at two times the halo virial radius). The
    green and the orange curves display the analytical halo model
    prediction summing the contribution of eq. (\ref{pkhalo1}) and
    (\ref{pkhalo2}), without normalising the 2-halo term, for two
    different minimum halo masses. \label{fighmpinocchio}}
\end{figure}

\begin{figure*}
  \includegraphics[width=0.49\hsize]{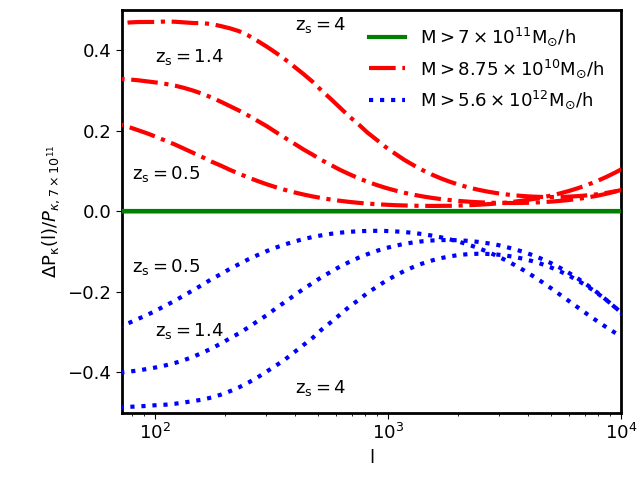}
  \includegraphics[width=0.49\hsize]{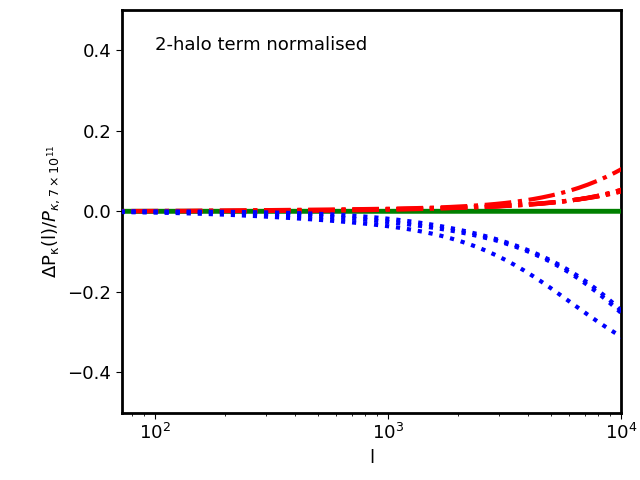}  
  \caption{Relative difference of the convergence power spectra,
    computed using the halo model at three different fixed source
    redshifts, using different minimum halo mass threshold in the
    integrals. Left panel shows the relative difference with respect
    to the prediction of the mass resolution of the reference run of
    the 1-halo and 2-halo terms summing the contributions from
    eq. (\ref{eqborn}) of the expressions in eq. (\ref{pkhalo1}) and
    (\ref{pkhalo2}). Right panel displays the relative difference of
    the total contributions where the 2-halo term has been normalised
    by eq. (\ref{pkhalo20}).\label{figh}}
\end{figure*}

In Fig.~\ref{fighmpinocchio} we show the halo model predictions for
the convergence power spectra, at three different source redshifts
(from bottom to top, $z=0.5$, $1.4$ and $4$) integrating the matter
power spectra as in eq. (\ref{eqborn}). The black dashed curves show
the prediction of the halo model summing the 1 and 2-halo term, with
the latter normalised by the eq. (\ref{pkhalo20}) to match on large
scales the linear theory power spectra (dotted light-blue lines). The
magenta dot-dashed curves exhibit the average convergence power
spectra of $512$ maps constructed by {\sc wl-moka} using haloes from
the {\sc pinocchio} reference runs; for comparison, the blue
dot-dashed curves display the results from the $512$ maps created
truncating the halo density profiles at two times the virial
radius. Green and orange curves display the prediction of the 1 plus
2-halo term not normalised. The fact that the trend of those curves is
similar as the {\sc pinocchio} plus {\sc wl-moka} indicates that the
convergence maps we have constructed using only haloes do not match
the predictions from linear theory on large scale. In the analytic
halo model formalism this is compensated only in the 2-halo term2,
normalising it by an effective bias contribution as in
eq. (\ref{pkhalo20}), without interfering on the matter density
distribution on small scales as described by the 1-halo term. However,
it is worth noticing that in the convergence maps constructed using
{\sc wl-moka} we cannot separate a priori the two terms; when
including the unresolved matter contribution using linear theory, this
will include somehow a small correlation between large and small
scales compensating for re-scaling the amplitude as in
eq. (\ref{eqampl}).

In Fig.~\ref{figh} we display the relative differences, with respect
to the case of $M_{min}= 7\times 10^{13}M_{\odot}/h$, of the total
analytical halo model convergence power spectra, at the same three
fixed source redshifts, assuming various minimum halo masses. While on
the left panel the 2-halo term is not normalised, in the right panel
eq. (\ref{pkhalo2}) has been normalised by the effective bias term as
in eq. (\ref{pkhalo20}). In this latter case we can notice that the
mass resolution of the halo model integrals has a negligible effect on
large scales, where the 2-halo term is forced to follow linear theory,
while on small scales (large $l$) it manifests in appreciable
differences.

\section{Results}
\label{results}

\begin{figure}
  \includegraphics[width=\hsize]{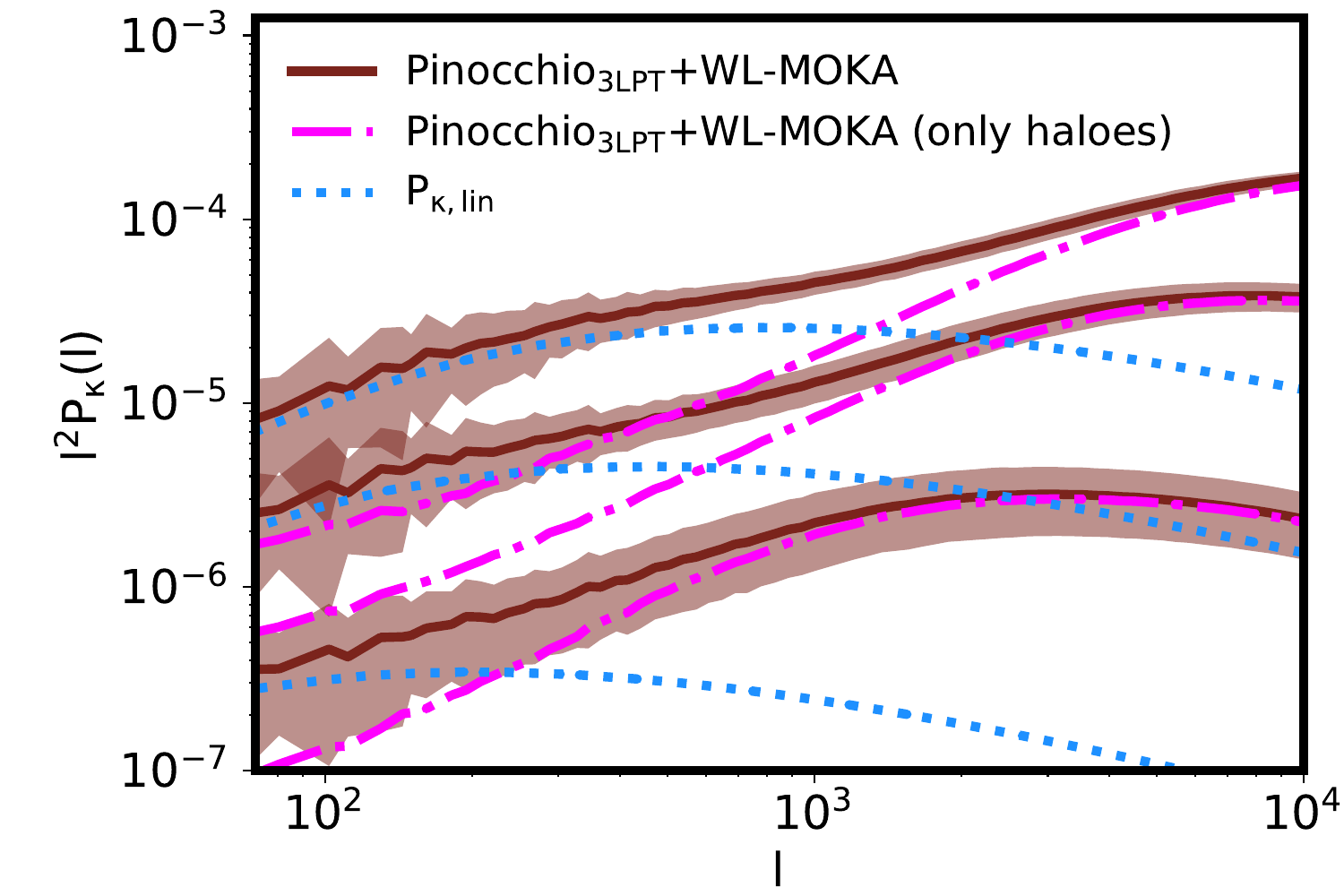}
  \caption{Convergence power spectra at three different source
    redshifts: from top to bottom $z_s=4$, $1.4$ and $0.5$. The dotted
    blue curves displays the predictions using linear power spectrum,
    dot-dashed purple the average halo contribution of 512 light-cones
    from \textsc{pinocchio}.  The red lines are the average
    measurements of 512 realisations of the convergence field from
    haloes and diffuse component, the shaded regions enclose the rms
    of the various realisations.\label{figpklhaloes}}
\end{figure}

\begin{figure}
  \includegraphics[width=\hsize]{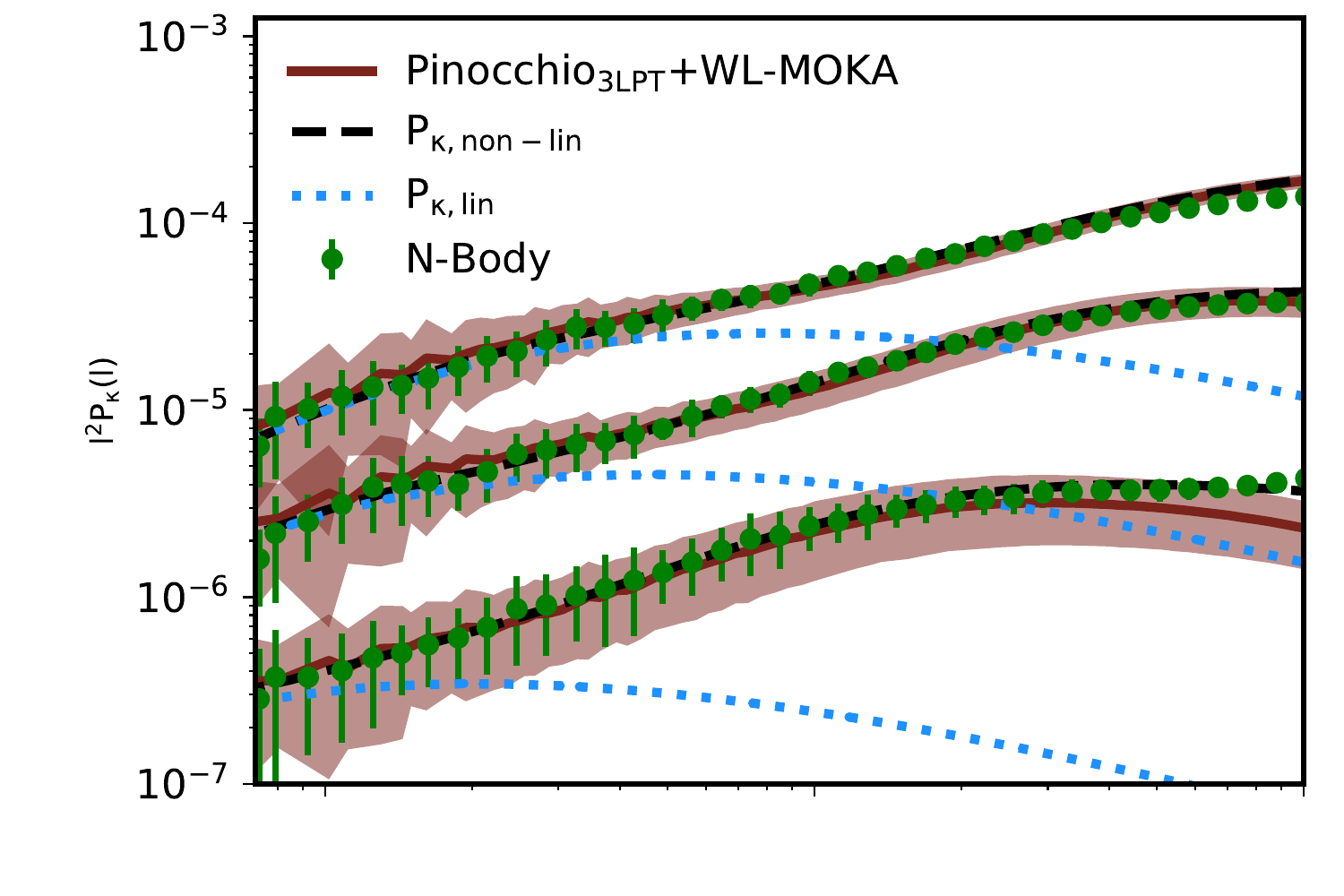}
  \vspace{-0.95cm} \\  
  \includegraphics[width=\hsize]{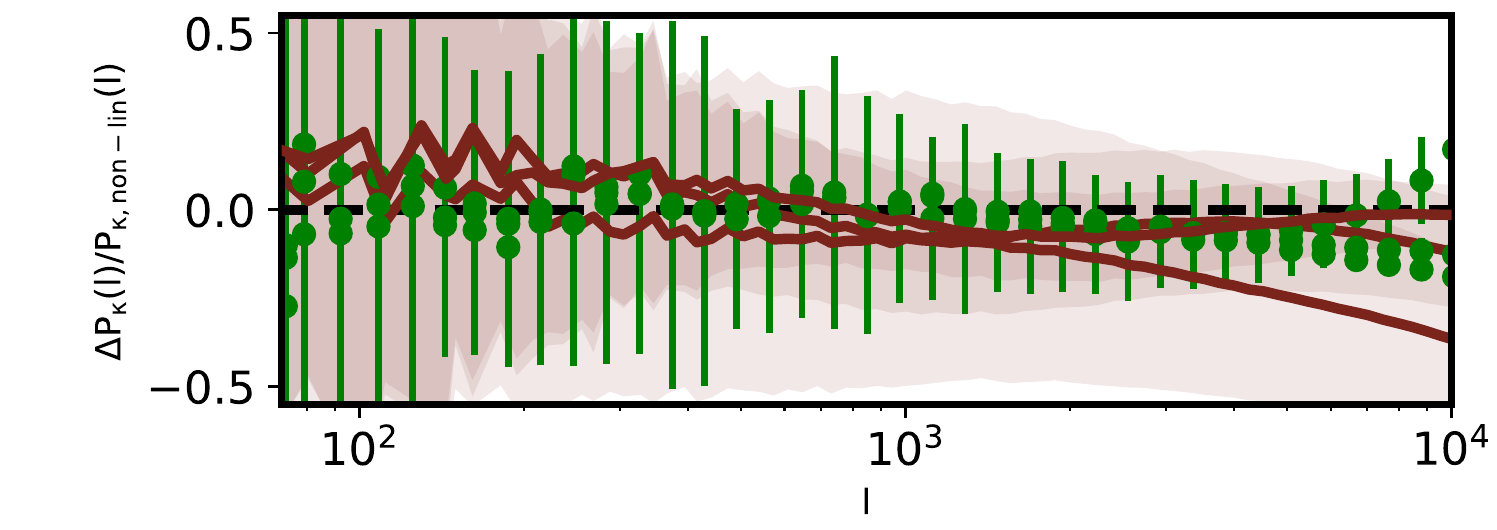}  
  \caption{Convergence power spectra at three different source
    redshifts, from top to bottom $z_s=4$, $1.4$ and $0.5$. We compare
    the predictions using approximate methods \textsc{pinocchio} and
    \textsc{WL-MOKA} (solid red curves) with measurements from
    light-cones extracted from a cosmological N-Body simulation using
    \textsc{MapSim}. Black dashed and blue dotted curves display the
    predictions using linear and non-linear power spectra from
    \textsc{CAMB}, the latter refers to the implementation by
    \citet{takahashi12}. In the bottom sub-panel we display the
    relative differences of the N-Body and the approximate convergence
    power spectra with respect to the non-linear predictions.
    \label{figpk}}
\end{figure}

In Fig.  \ref{figpklhaloes} we show the convergence power spectra at
three source redshifts $z_s=4$, $1.4$ and $0.5$ from top to bottom,
respectively. The dotted blue curves represent the predictions from
linear theory using eq.~(\ref{eqborn}), the dot-dashed magenta ones
the average convergence power spectra of the haloes over 512 different
realisations of {\sc pinocchio} light-cones -- left panel of
Fig.~\ref{figmap1}.  The red (in particular this is a falu red) solid
curves display the average power spectra of the 512 simulated
convergence fields in which we include also the contribution from
unresolved matter (central panel of Fig.~\ref{figmap1}), the shaded
regions enclose the standard deviation of the various realisations.
We remind the reader that while the numerical cosmological simulation
has been run with only one initial condition displacement field, all
$25$ light-cones generated from this have been constructed by
randomising the simulation box by translating the particles and
redefining the simulation centre when building-up the light-cones.  In
Fig.~\ref{figpk} we compare the convergence power spectra of our final
maps (red solid curves) with the ones from the N-body simulation
(green dots). From the figure we notice a very good agreement between
our model and the N-body results up to $l\simeq 3\times 10^3$ where
the green data points start to deviate at low redshifts due to
particle shot-noise \citep{giocoli16a}, and the model tends to move
down because of the absence of concentrated low mass haloes below the
numerical resolution. The black dashed curves show the corresponding
convergence power spectra obtained by integrating the non-linear
matter spectra implemented in \textsc{CAMB} by \citet{takahashi12}.
In the bottom sub-panel we display the relative difference between the
N-Body predictions and the approximate methods with respect to the
non-linear projected matter power spectrum.  From the figure we can
see that the non-linear predictions are recovered with percent
accuracy from $l=72$ (which is the angular mode corresponding to 5
deg) to $l=3\times 10^3$ (the largest mode expected to be covered by
future wide field surveys) using our approximate methods.

\begin{figure*}
  \includegraphics[width=0.7\hsize]{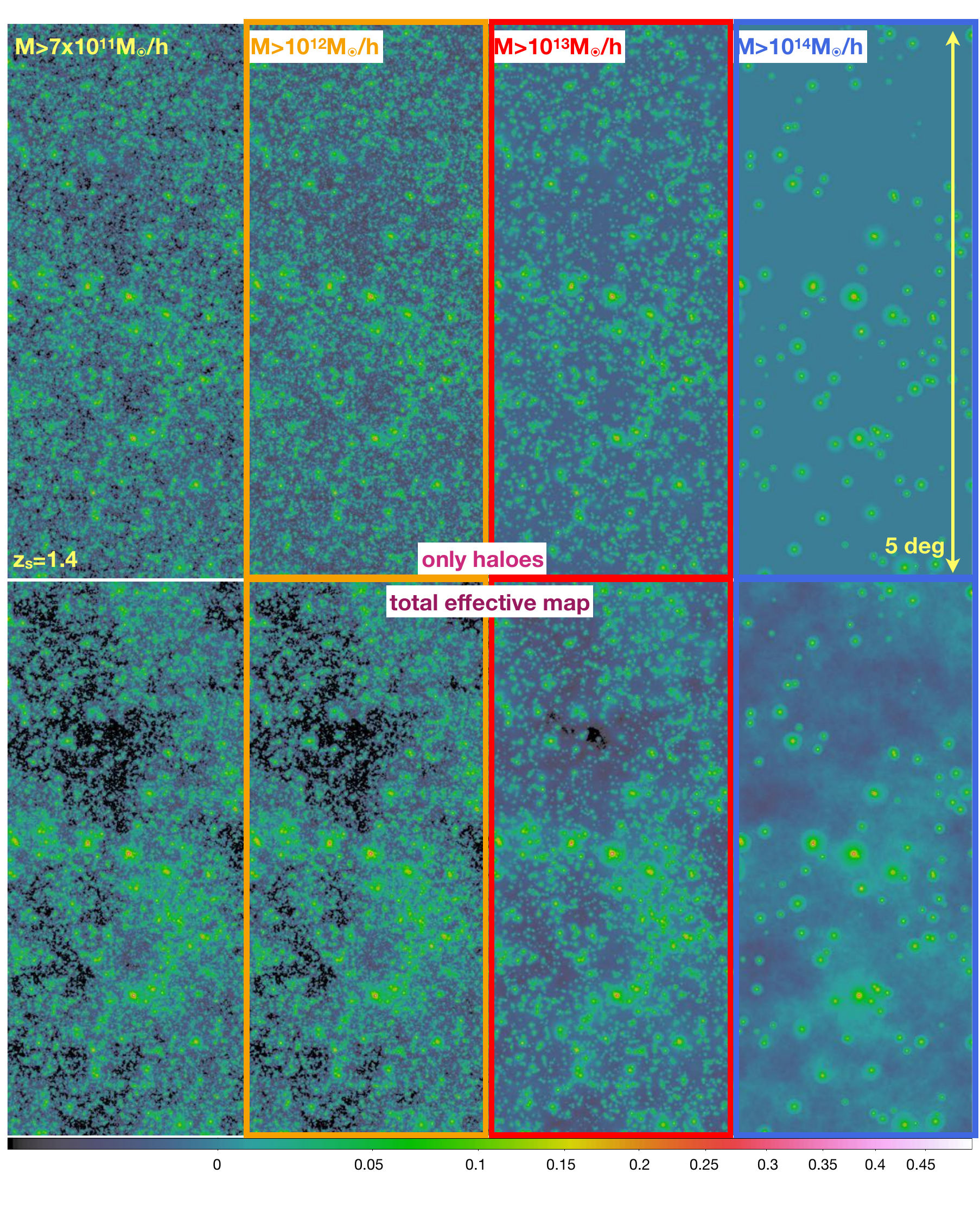}
  \caption{Convergence maps for $z_s=4$ constructed adopting different
    minimum mass thresholds: we use all haloes with mass larger than
    $7 \times 10^{11}$, $10^{12}$, $10^{13}$ and
    $10^{14}\;M_{\odot}/h$ from left to right, respectively.  The top
    panels show the halo contribution while the bottom ones include
    also the linear contribution coming from unresolved matter, below
    the considered minimum halo mass.\label{mapresolution}}
\end{figure*}

\begin{figure*}
  \includegraphics[width=0.33\hsize]{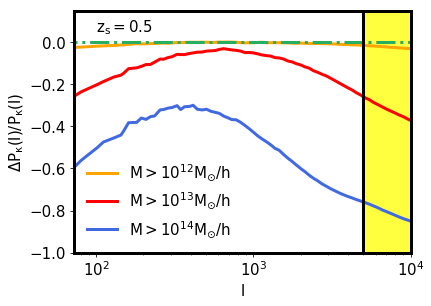}
  \includegraphics[width=0.33\hsize]{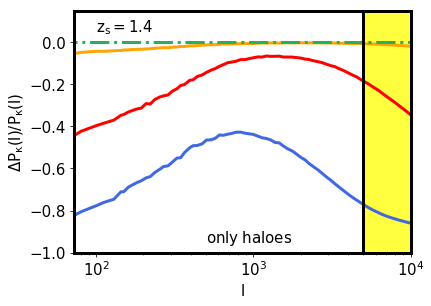}
  \includegraphics[width=0.33\hsize]{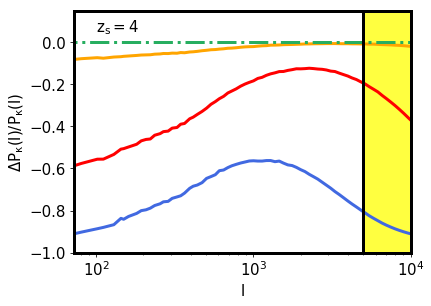}
  \includegraphics[width=0.33\hsize]{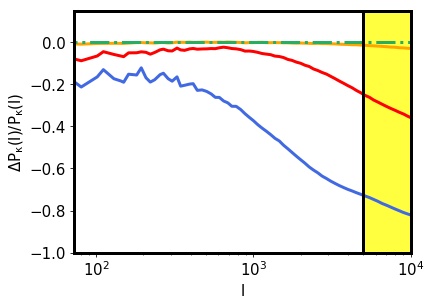}
  \includegraphics[width=0.33\hsize]{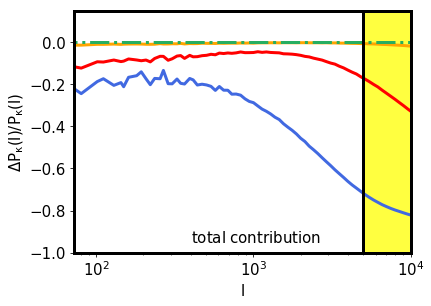}
  \includegraphics[width=0.33\hsize]{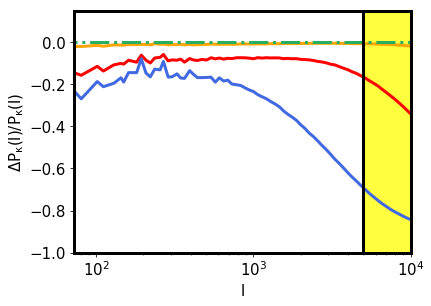}
  \caption{\label{kapparesolution}Relative difference in the
    convergence power spectra adopting various minimum halo mass
    thresholds at three different source redshifts: $z_s=0.5$, $1.4$
    and $4$ from left to right, respectively.  Orange, red and blue
    curves display the measurements considering minimum masses of
    $10^{12}$, $10^{13}$ and $10^{14}\;M_{\odot}/h$ with respect to
    the reference case that has a minimum halo mass of $7 \times
    10^{11}\;M_{\odot}/h$.  As in Fig.  \ref{mapresolution}, the top
    panels show the average contribution of $512$ realisations
    constructed using only haloes while the bottom ones show the
    average including also the linear contribution coming from
    unresolved matter, below the considered minimum halo mass. The
    yellow shaded region marks the region of $l > 5\times
    10^3\;\mathrm{rad}$, up to which future wide field surveys are
    expected to measure the weak lensing convergence power spectra
    \citep{euclidredbook}.}
\end{figure*}

\subsection{Stability with minimum halo mass}

We profit from the large statistic samples available from the
\textsc{pinocchio} runs and we investigate how the reconstruction
\citep{giocoli17} of large-scale projected power missing from haloes
behaves, as a function of redshift, when a higher threshold for
minimum halo mass is used.  We run our {\sc wl-moka} pipeline adopting
different threshold for the minimum halo mass.  In
Fig.~\ref{mapresolution} we display the convergence map for $z_s=1.4$
constructed with a minimum halo mass of $7\times 10^{11}$, $10^{12}$,
$10^{13}$ and $10^{14}\;M_{\odot}/h$ from left to right,
respectively. The value of $7\times 10^{11}\;M_{\odot}/h$ corresponds
to our minimum halo mass in the {\sc pinocchio}$_{\rm 3LPT}$ reference
run, while the case with minimum mass of $10^{14}\;M_{\odot}/h$ will
help us in better understanding the contribution of galaxy cluster
size-haloes to the weak lensing signal. In the figure, top and bottom
panels display the halo and the total contributions to the
convergence, respectively. In the latter case we add the linear matter
density contribution due to unresolved matter below the corresponding
mass threshold limit. From the figure we notice that while clusters
represent the high density peaks of the convergence field, smaller
mass haloes trace the filamentary structure of the matter density
distribution and contribute ,in projection, to increasing the lensing
signal \citep{martinet17,shan17,giocoli18a}.

In Fig.~\ref{kapparesolution} we present the relative difference of
the average convergence power spectra, for 512 different realisations,
as computed from the different mass threshold maps with respect to the
reference one, that has a mass resolution of $7\times
10^{11}\;M_{\odot}/h$.  Panels from left to right consider different
fixed source redshift $z_s=0.5$, $1.4$ and $4$, respectively; while on
the top we show only the halo contributions.  In the bottom we account
also for the contribution from unresolved matter. These figures help
us to quantify the contribution of various haloes to the convergence
power spectra. In particular, from the top panels we can notice that
the contribution of clusters evolves with the redshift going from an
average value of approximately $40\%$ for sources at $z_s=0.5$ to
$10\%$ for $z_s=4$. This is an effect where the cluster contribution
is modulated by the lensing kernel, depending on the considered source
redshift, and it is sensitive to the redshift evolution of the cluster
mass function. On the other side, in the bottom panels we can notice
that the average power spectra from total effective convergence maps
built using only clusters deviates by approximately $30\%$ with
respect to the reference ones, independently of the source redshift.
The yellow shaded region, in the right part of each panel, indicates
where future wide field surveys like the ESA Euclid mission will not
be able to provide any reliable data; future wide field missions will
be able to probe the convergence power spectrum up to $l\approx 3
\times 10^3$.

As discussed by \citet{munari17} and \citet{paranjape13} in {\sc
  pinocchio} the bias of dark matter haloes is a prediction. This
applies also when the code is interfaced with {\sc wl-moka} in
creating effective convergence maps. In particular the square-root of
the ratio at large angular scales (the comoving scale of $k = 0.1 h$
Mpc$^{-1}$ corresponds approximately to $l\approx 150$, $300$ and
$520$ for source redshift $z_s=0.5$, $1,4$ and $4$, respectively)
between the halo and the total convergence power spectrum gives us a
measure of the projected halo bias. This can be modelled using the two
halo term of the projected halo model as discussed in
Sec.~\ref{hmsection}.

\begin{figure*}
  \includegraphics[width=0.33\hsize]{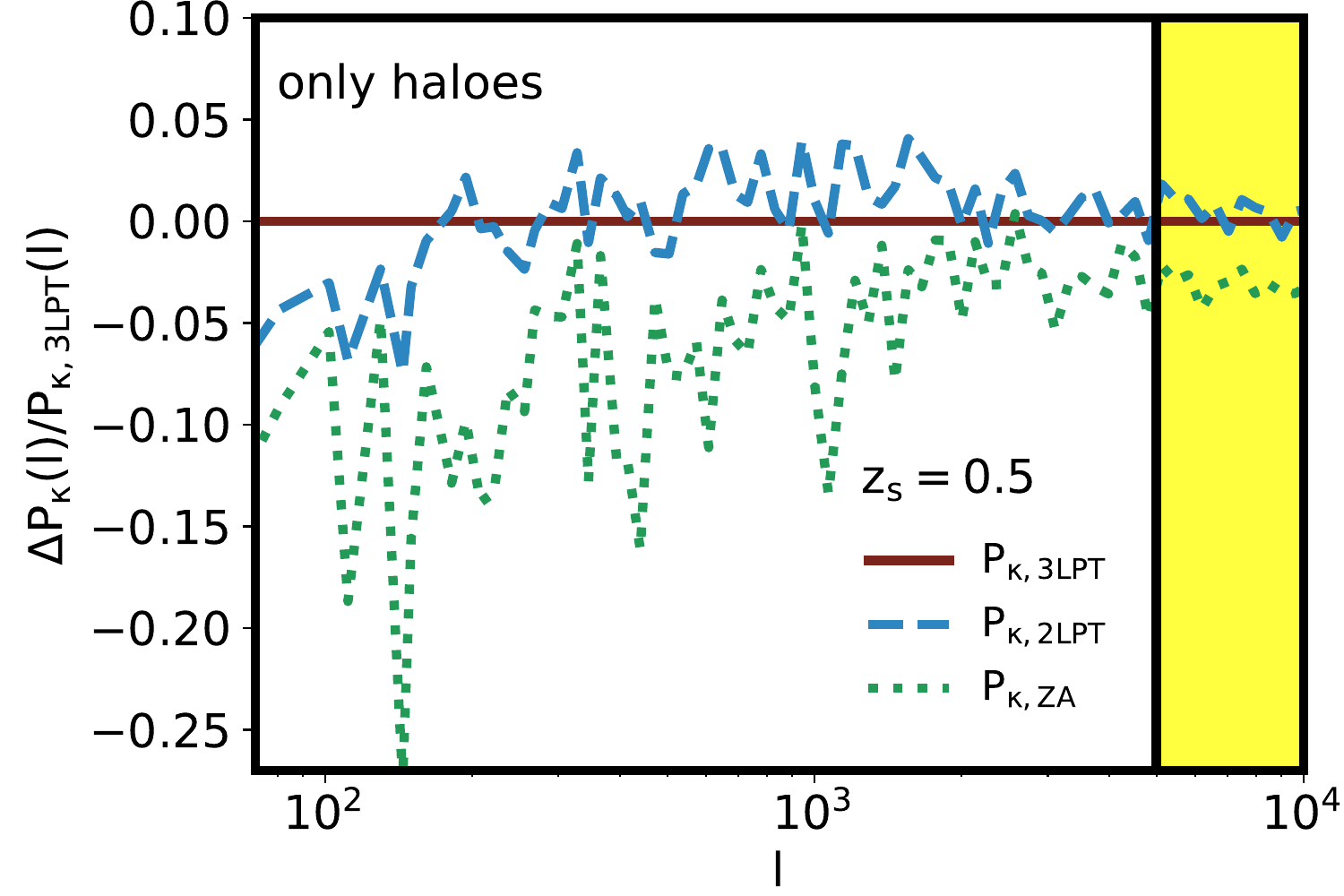}
  \includegraphics[width=0.33\hsize]{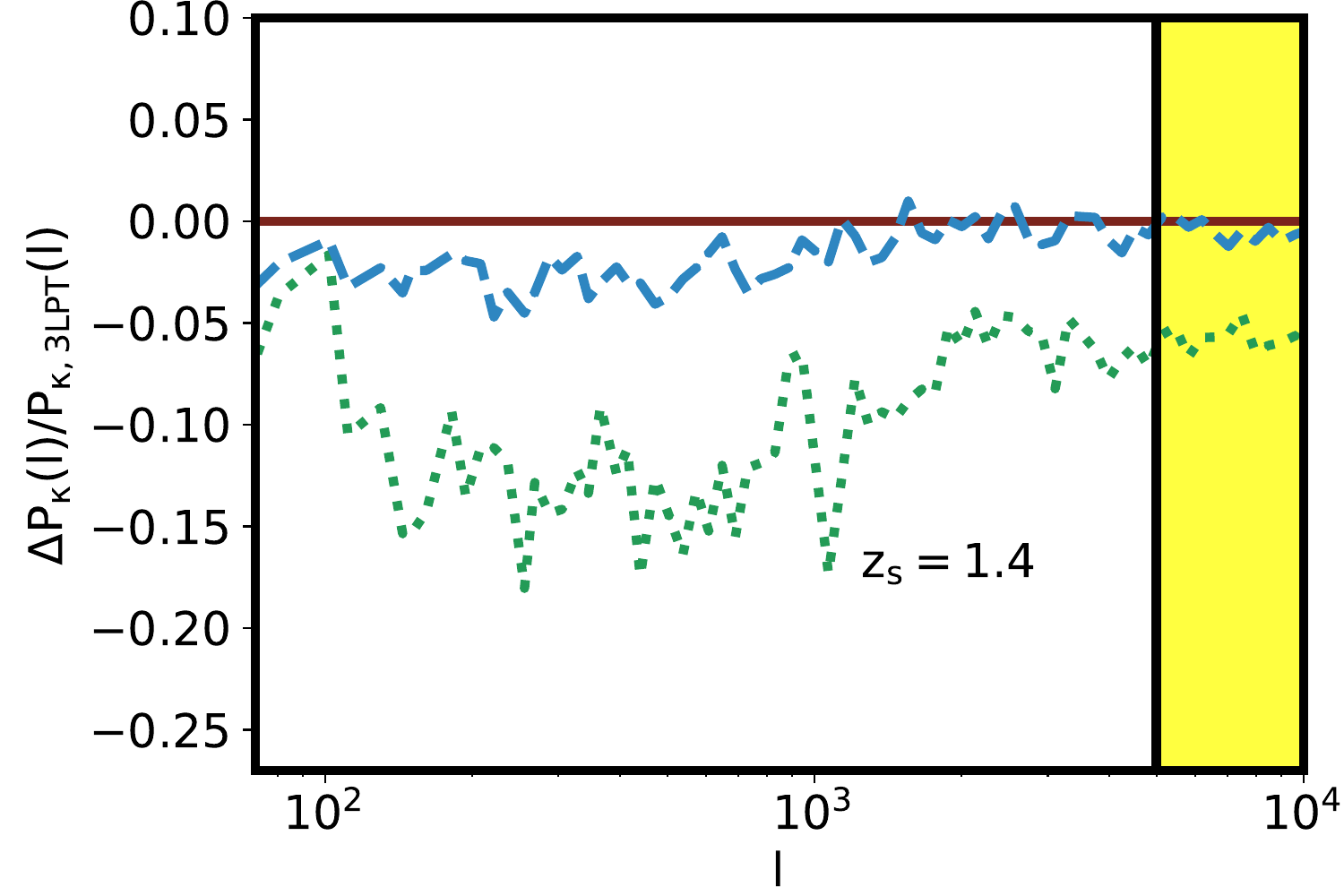}
  \includegraphics[width=0.33\hsize]{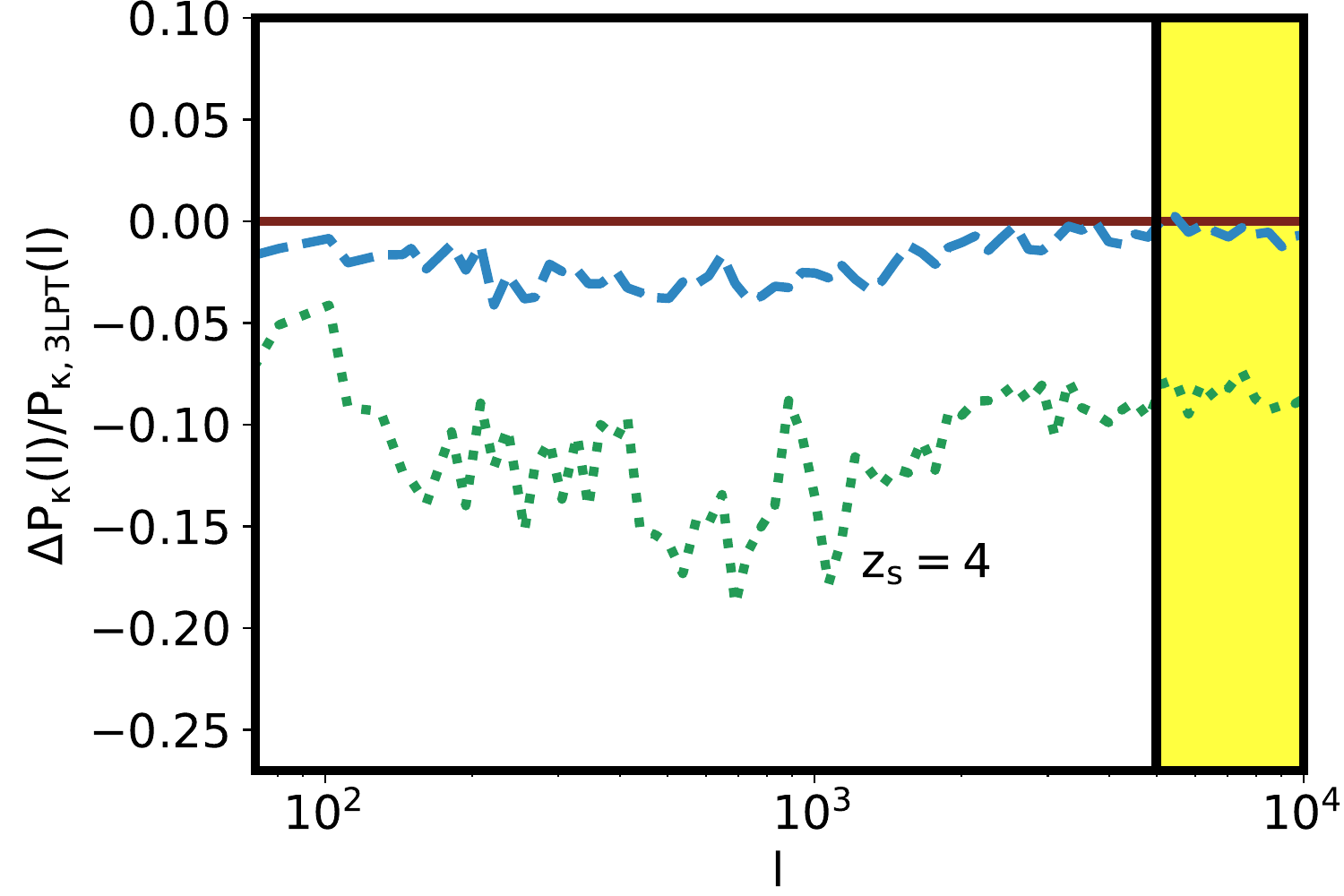}
  \includegraphics[width=0.33\hsize]{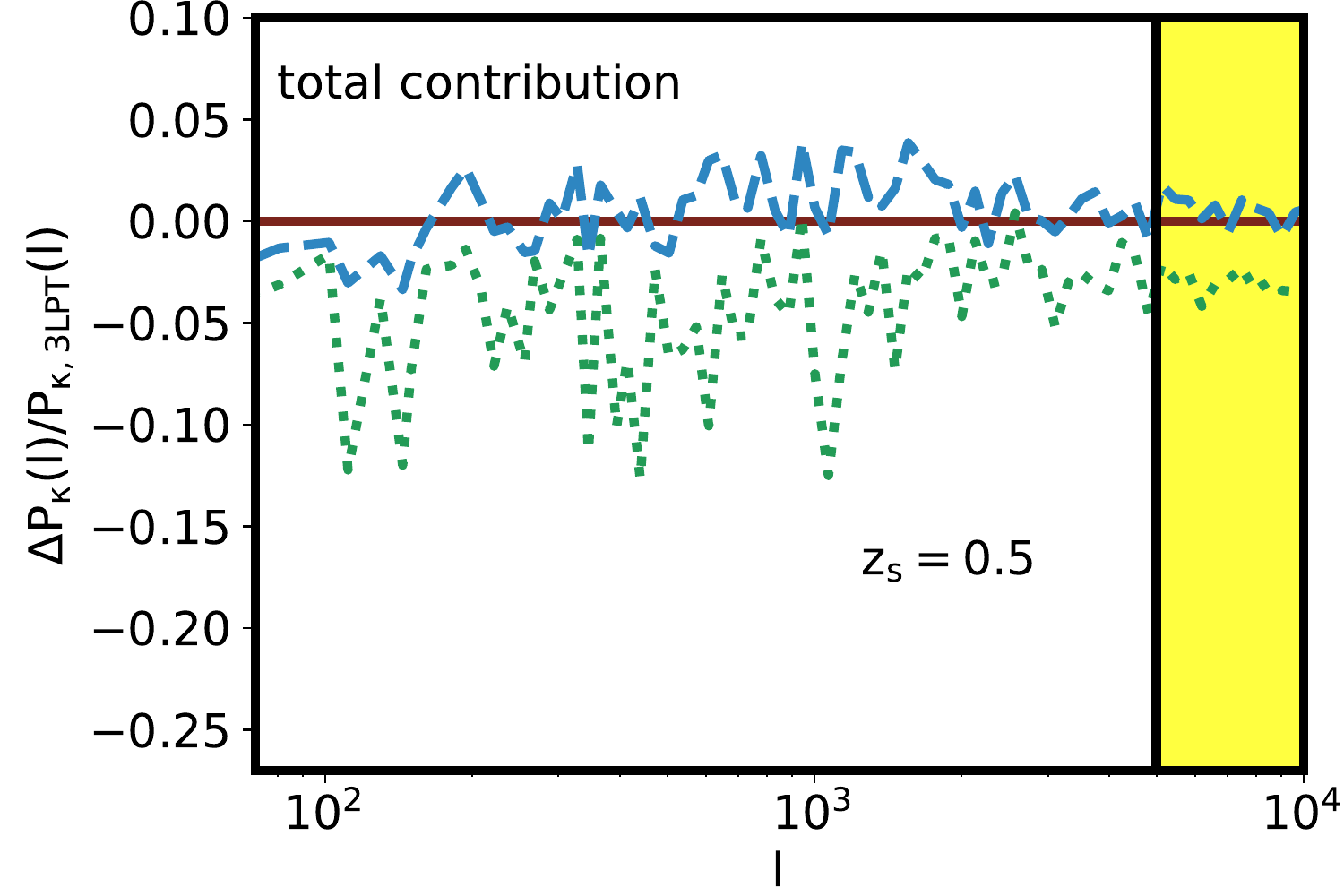}
  \includegraphics[width=0.33\hsize]{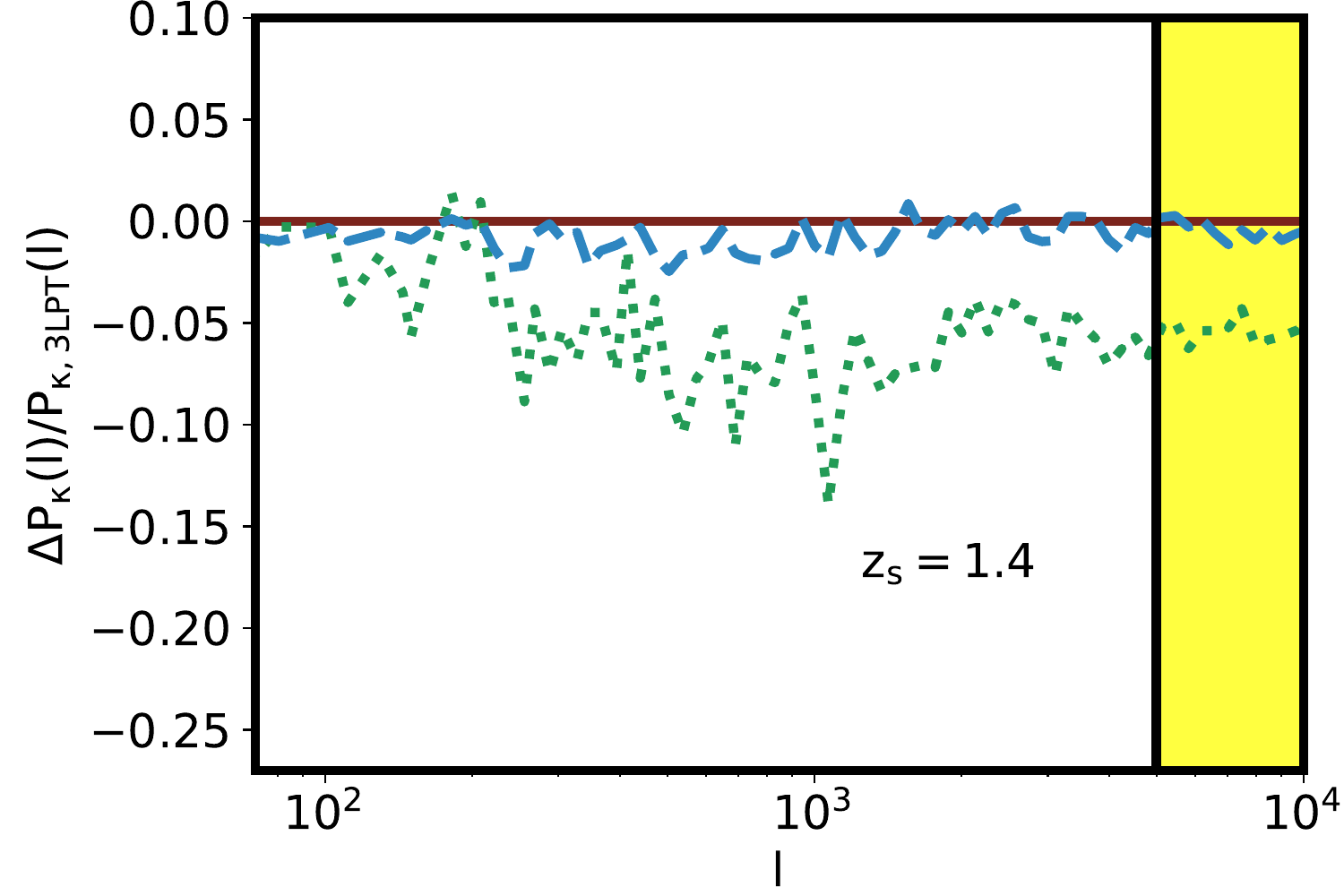}
  \includegraphics[width=0.33\hsize]{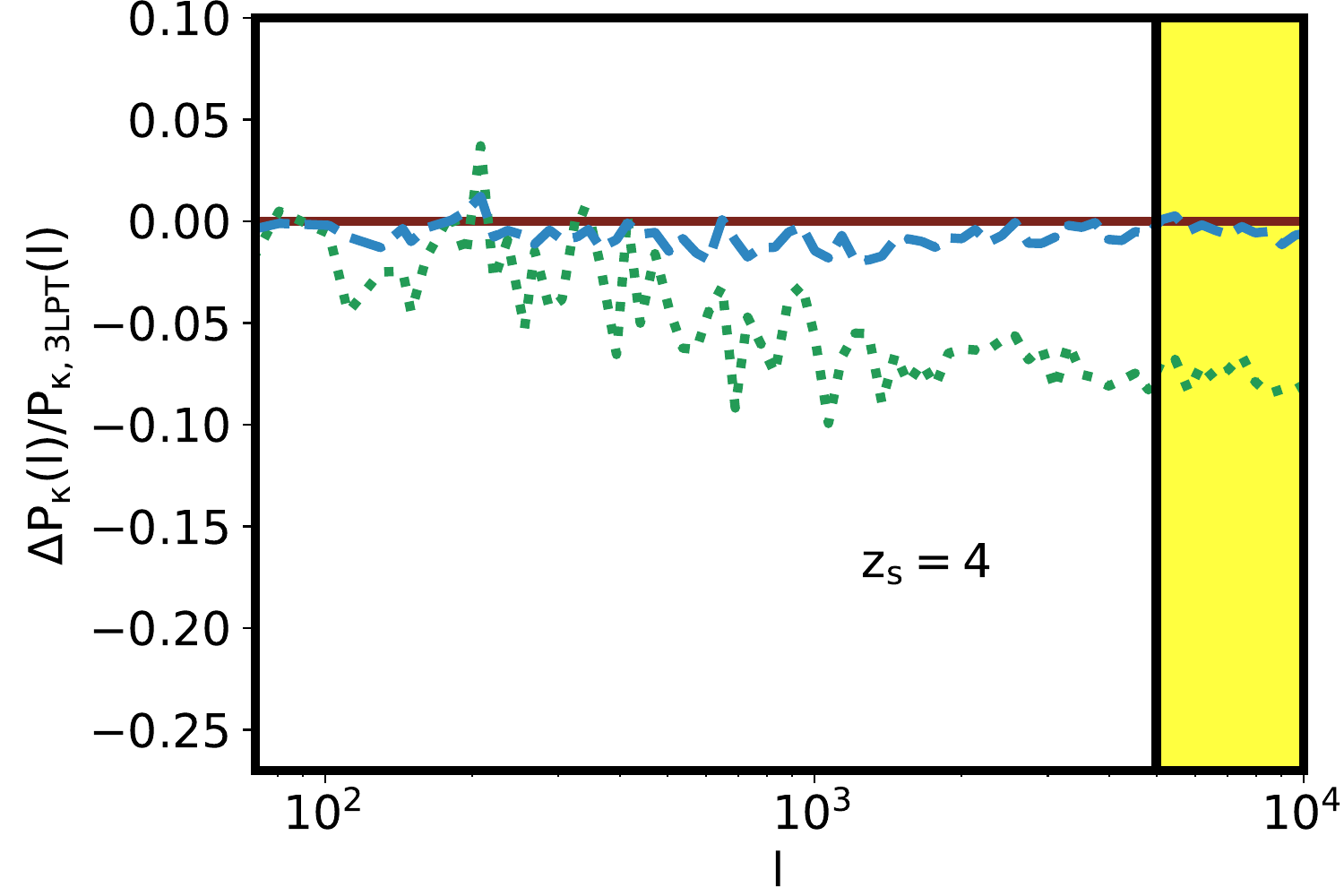}
  \caption{Relative convergence power spectra between maps constructed
    using \textsc{pinocchio} 2LPT (dashed blue) and ZA (dotted green)
    with respect to 3LPT. From left to right we show the cases for the
    three considered source redshifts: $z_s=0.5$, $1.4$ and $4$,
    respectively. Top and bottom panels show the cases in which the
    maps are constructed using only haloes or haloes plus diffuse
    matter.\label{ZS2LPT3LPTfig}}
\end{figure*}

\subsection{LPT order for the displacement fields}

As presented by \citet{munari17}, in the new version of {\sc
  pinocchio} the user can run the code adopting different LPT orders
for particle displacement fields: ZA (Zel'dovich Approximation), 2LPT
and 3LPT. In this section we investigate how the reconstructed
convergence power spectra, at different fixed source redshifts,
depends on the adopted displacement order. In Fig.~\ref{ZS2LPT3LPTfig}
we show the relative difference of the convergence power spectra at
three source redshifts. Blue dashed and green dotted curves display
the relative difference of the ZA and 2LPT cases with respect to the
3LPT.  While the top panels exhibit the power spectra due only to
resolved haloes in the light-cone simulations, the bottom ones account
also for the corresponding contribution due to matter among haloes.
In the bottom panels, where we have the total contribution, while the
2LPT case differs from the 3LPT case by less than $1\%$, the ZA case
tends to deviate more than $5-7\%$ with larger deviations for larger
angular modes (small scales) reaching values of $\sim 10\%$.  We
remind the reader that the average measurements are done only on $25$
different light-cone runs and the simulations, with various
displacement prescription, share the same seeds and phases, when
generating the initial conditions.

\begin{figure*}
  \includegraphics[width=0.4\hsize]{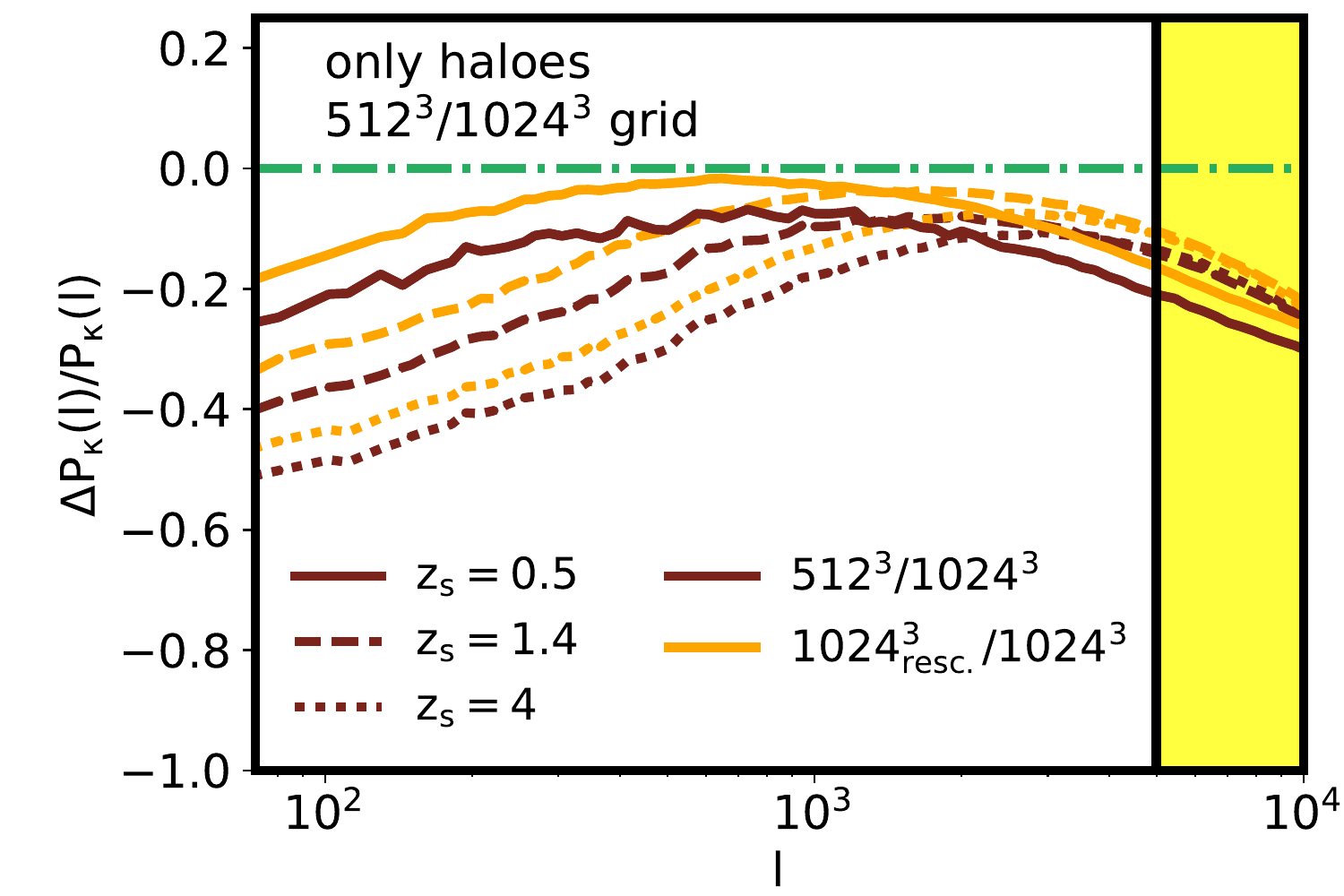}
  \includegraphics[width=0.4\hsize]{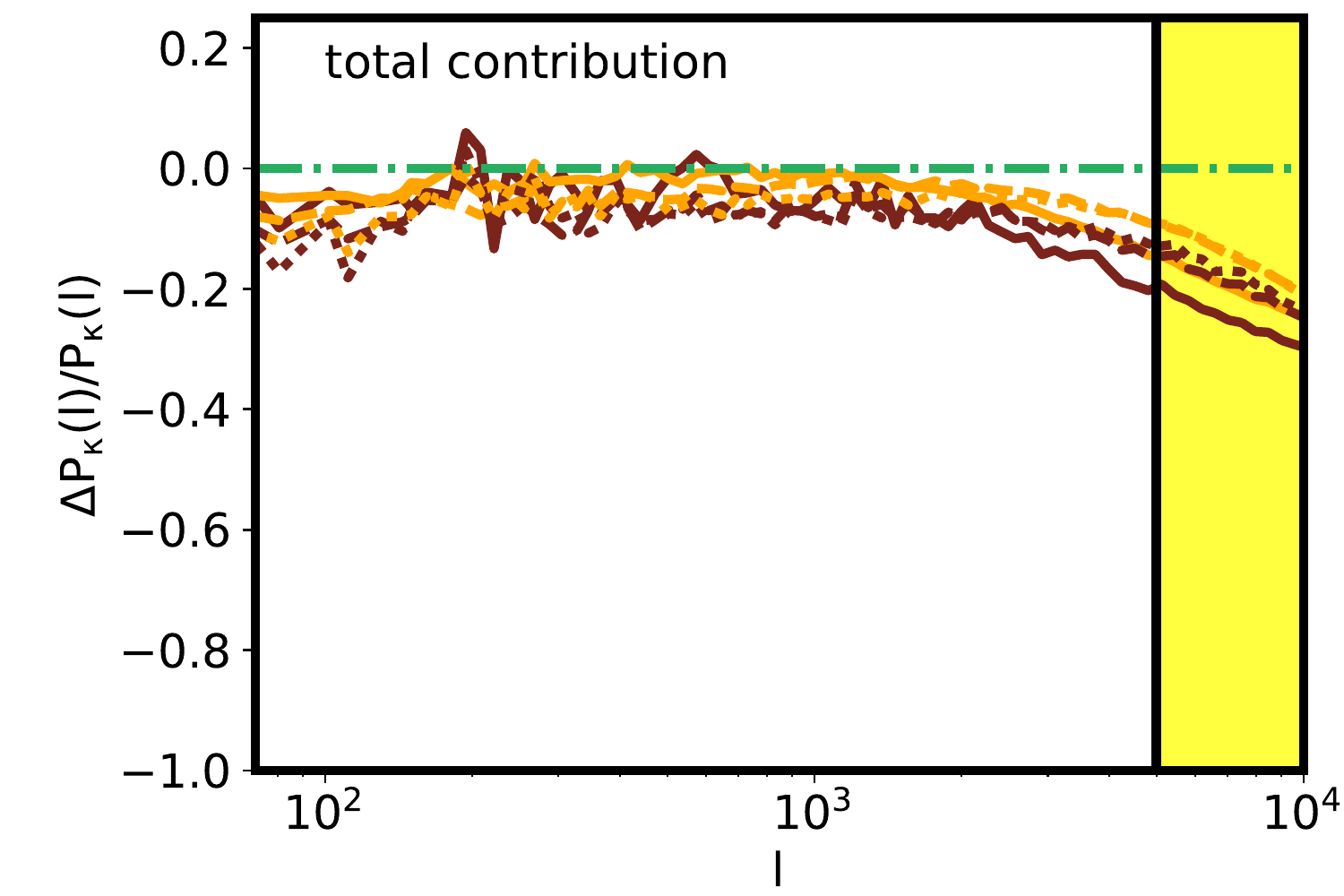}
  \caption{Relative difference between the $512^3$ and the $1024^3$
    simulations. The different curves show the results for the three
    source redshifts, left and right panels refer to the case where we
    consider the contribution only from haloes or we include also the
    diffuse matter using linear theory, respectively. The orange
    curves display the relative difference when building the maps from
    the $1024^3$ we adopt the same minimum halo mass resolution of the
    $512^3$, namely $5.6\times10^{12}M_{\odot}/h$.\label{1024512fig}}
\end{figure*}

\subsection{Resolution of Approximate Method Simulations}

The reconstructed 2D field using the halo model technique depends also
on the resolution of the grid mesh on the top of which the halo
catalogue is constructed in the PLC. As for numerical simulations that
solve the full N-Body relations, {\sc pinocchio} runs are much faster
when the displacement grid is coarser. In Fig. \ref{1024512fig} we
show the case in which we compare the convergence power spectra of
$512$ approximate methods past-light-cones with grid resolution of
$512^3$ and our reference run of $1024^3$; in both cases we assume a
3LPT displacement field and the same initial seeds for the initial
conditions. The orange curves display the relative difference of the
$1024^{3}$ maps re-scaled to the same resolution of the $512^3$ with
respect to the reference resolution run.  Left and right panel show
the comparison of the power spectra coming only from the haloes and
from the full projected matter density distribution (haloes plus
unresolved matter treated using linear theory). From the left figure
we can see that the runs with lower resolution have a redshift
dependence trend at low angular modes (larger scales). These are
integrated effects due to the haloes that mainly contribute to the
lensing signal when accounting for the lensing kernel at a given
source redshift. However, from the right figure we can notice that
when adding the contribution coming from unresolved matter using
linear theory (right panel) this redshift tendency disappears, and on
average up to $l=3\times 10^3$ the relative difference of the low
resolution run is between 5 and $10\%$. At those scales it is evident
that the simulations constructed from the coarser grid are missing
concentrated small haloes. In re-scaling the $1024^3$ to the same map
resolution of the $512^3$, namely $5.6\times10^{12}\,M_{\odot}/h$, we
notice few percent deviations probably attributed to the different
displacement grids of the two runs, halo finding threshold resolution
when linking particles in friends of friends and halo bias that enter
on the two halo term on large scale, even if the two PLC have the same
minimum halo mass.

\begin{figure}
  \includegraphics[width=\hsize]{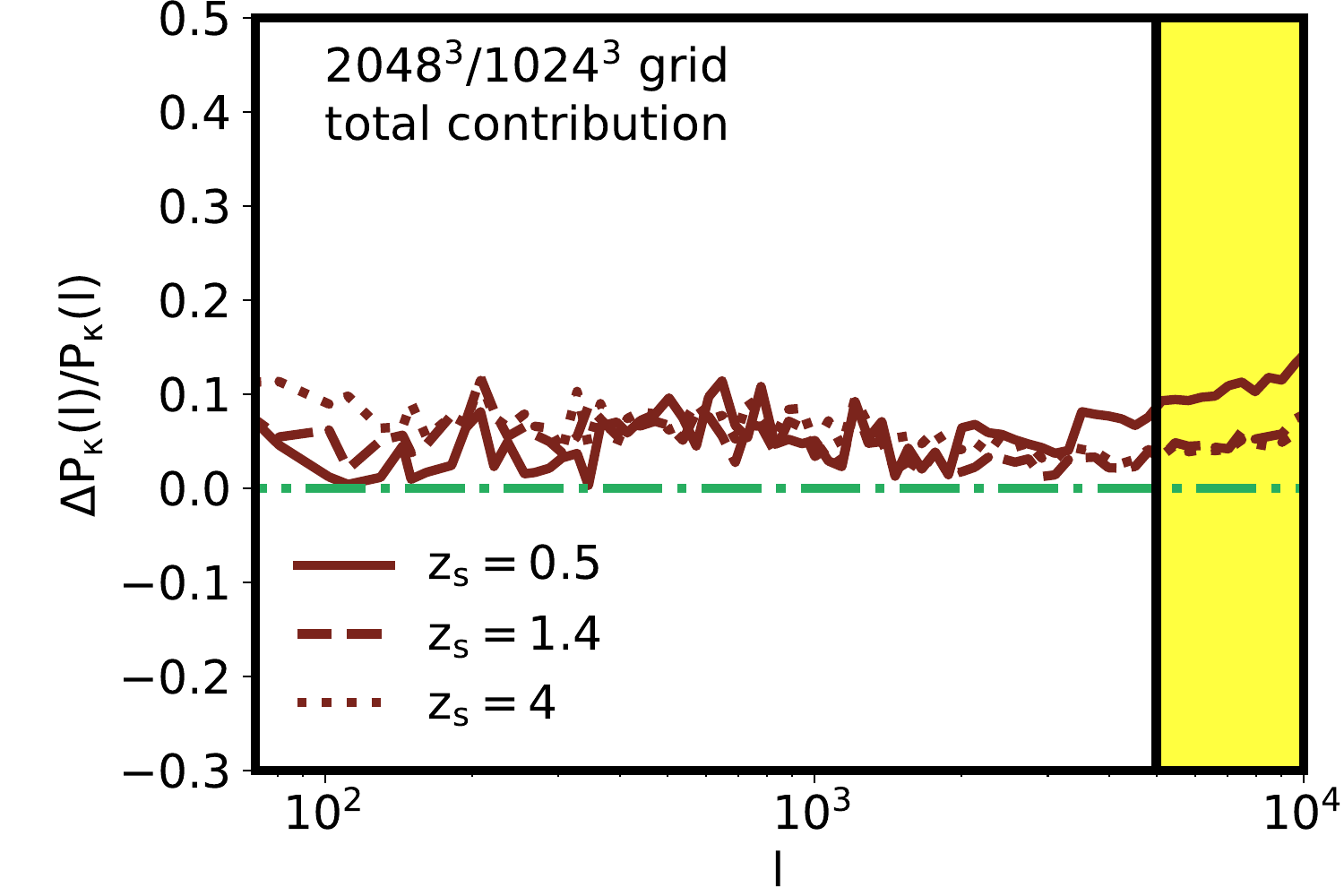}
  \caption{Relative difference between the convergence power spectra
    in the $2048^3$ and $1024^3$ runs.  The relative difference has
    been computed as average of $25$ difference past-light-cone
    realisations considering both haloes and diffuse
    matter.\label{10242048fig}}
\end{figure}

Fig. \ref{10242048fig} displays the relative difference between full
power spectra of $25$ high resolution and reference runs that share
the same initial condition seeds. As in the previous figures, the
different line types refer to various fixed source redshifts. As we
can read from Table \ref{tabsims} the run with $2048^3$ grid mesh has
a minimum halo mass a factor of 8 lower than the one with
$1024^3$. The figure shows that convergence power spectra computed
from the high resolution runs are approximately $3-5\%$ higher that
the ones from our reference simulations. This difference probably
arises from the fact that when including the contribution from diffuse
matter in our convergence halo maps we are not able to specifically
separate the 1 and 2-halo term -- as we do analytically in the halo
model as discussed in section {\ref{hmsection}}, causing a small shift
up. However, we want to underline that the main goal of our approach
is to be able to build a fast and self-consistent method for
covariances; many faster and more accurate approaches already exist in
term of non-linear power spectra that will be used for modelling the
small scales observed signals from future weak lensing surveys
\citep{peacock96,smith03,takahashi12,debackere19,schneider19}.

For summary, we address the reader to Appendix \ref{appx2} where we
show the convergence maps build using different particle and grid
resolutions, while adopting the same initial displacement seed and
phases. The maps refer to light-cones build only from haloes up to
$z_s=1.4$.

\begin{figure}
  \includegraphics[width=1\hsize]{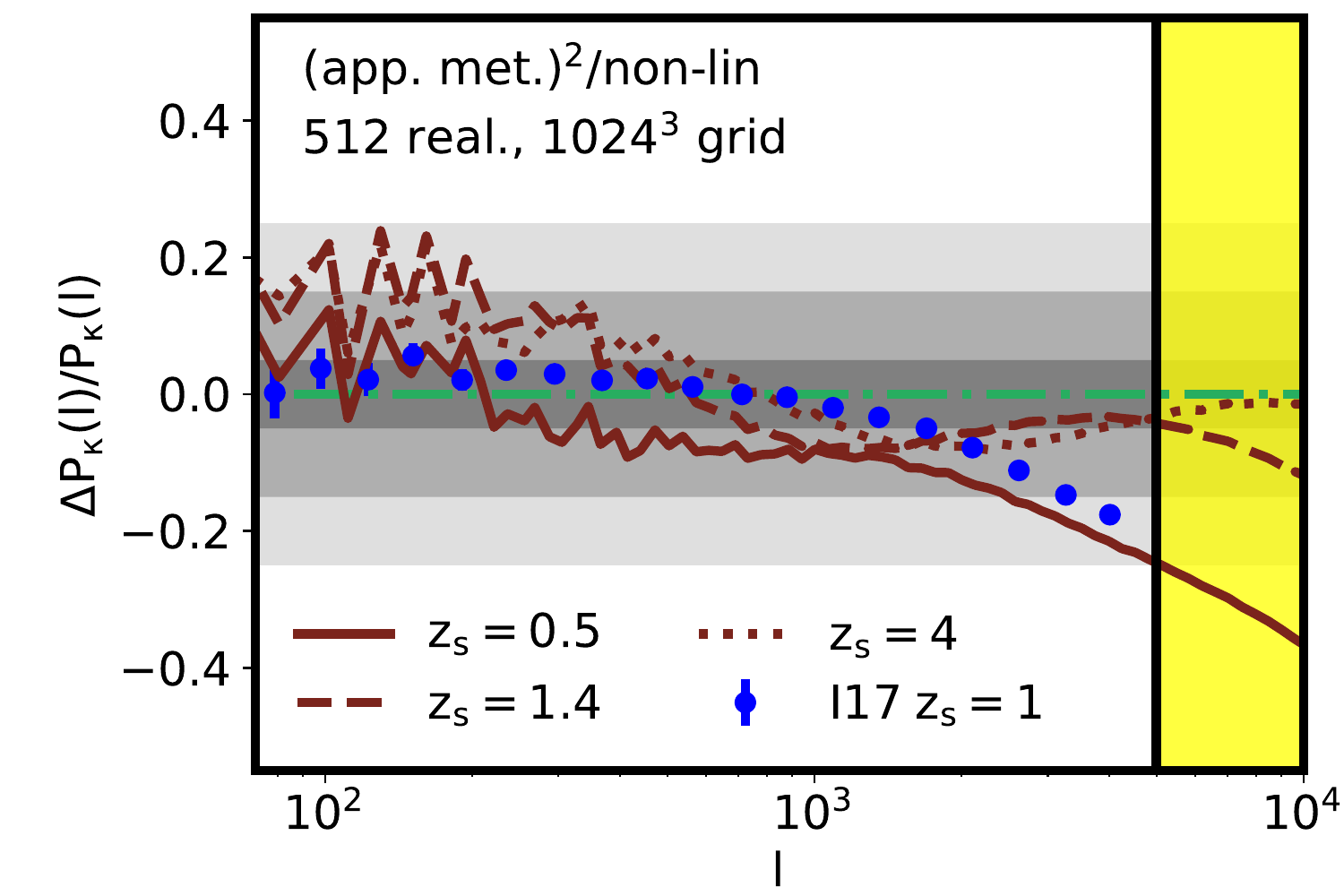}
  \caption{Relative difference between convergence power spectra
    computed using approximate methods and the prediction from
    non-linear matter power spectrum as implemented in {\sc CAMB} by
    \citet{takahashi12}. The different curves red curves show the
    results for the three considered source redshifts, while the data
    points displays the results obtained by \citet{izard18} for
    $z_s=1$ comparing their approximate methods for weak lensing,
    based on ICE-COLA, with respect to the reference MICE
    simulation. In the figure the red curves show the average values
    obtained on $512$ \textsc{pinocchio} light-cone realisations with
    a grid resolution of $1024^3$.  Dark-grey, grey and light-grey
    bands indicate $5$, $15$ and $25\%$ relative
    differences. \label{pkllast}}
\end{figure}

In Fig. \ref{pkllast} we show the relative difference of the average
power spectra computed from $512$ different past light-cones with
$1024^3$ grid mesh with respect to the prediction obtained integrating
the non-linear matter power spectrum from {\sc CAMB} using the
\citet{takahashi12} implementation. The dark grey, grey and light grey
regions mark the relative difference of $5$, $15$ and $25\%$, blue
data points with the corresponding error bars are the convergence
power spectra prediction obtained by \citet{izard18} using ICE-COLA
with respect to the MICE simulation for sources at $z_s=1$. These
comparisons strengthen the power of the two methods {\sc pinocchio}
plus {\sc wl-moka} in reconstructing the projected non-linear power
spectrum, and the flexibility to use different fixed source redshifts
or a defined source redshift distribution of sources. The limit of
these runs are related to the small field of view we have decided to
simulate, but this will be extended in a future dedicated work as well
as the possibility to construct convergence on
HEALPix\footnote{\href{https://healpix.jpl.nasa.gov}{https://healpix.jpl.nasa.gov}}
maps.

\begin{figure*}
\includegraphics[width=0.33\hsize]{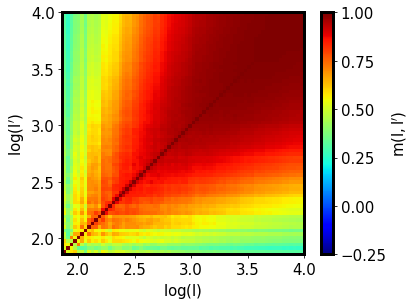}
\includegraphics[width=0.33\hsize]{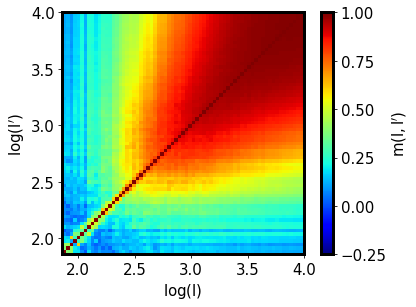}
\includegraphics[width=0.33\hsize]{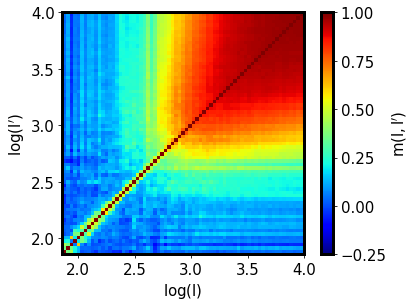}
\includegraphics[width=0.33\hsize]{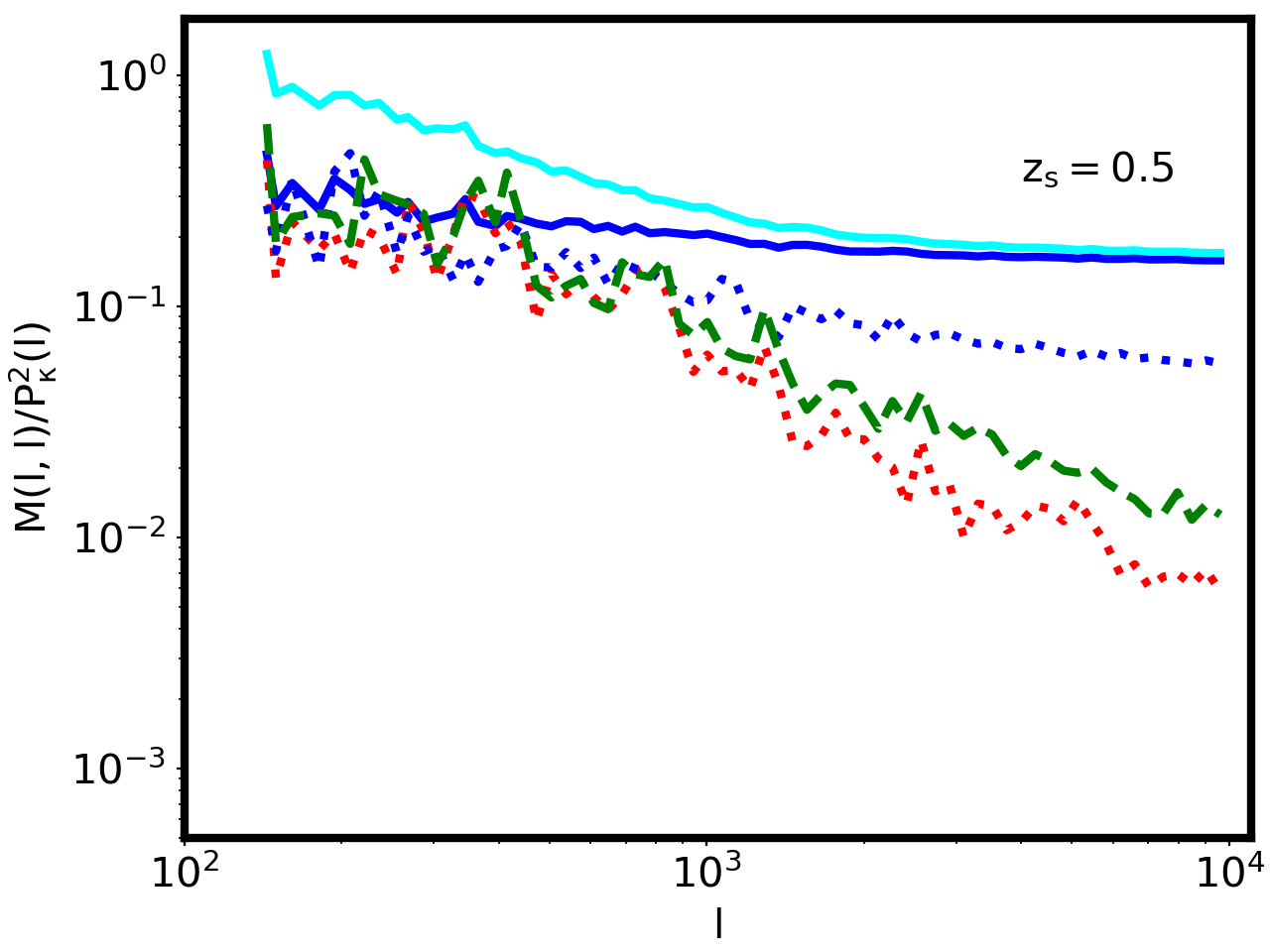}
\includegraphics[width=0.33\hsize]{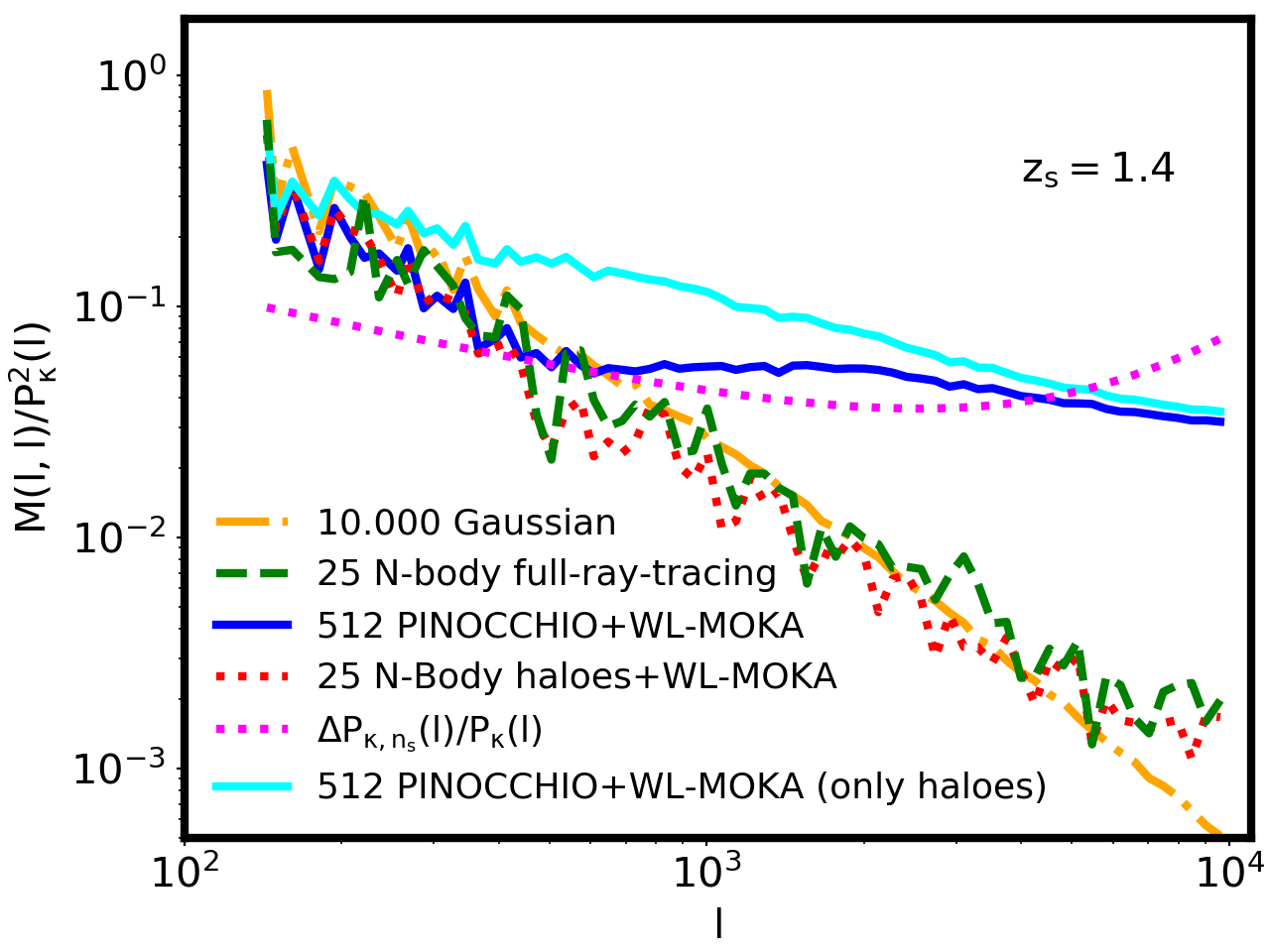}
\includegraphics[width=0.33\hsize]{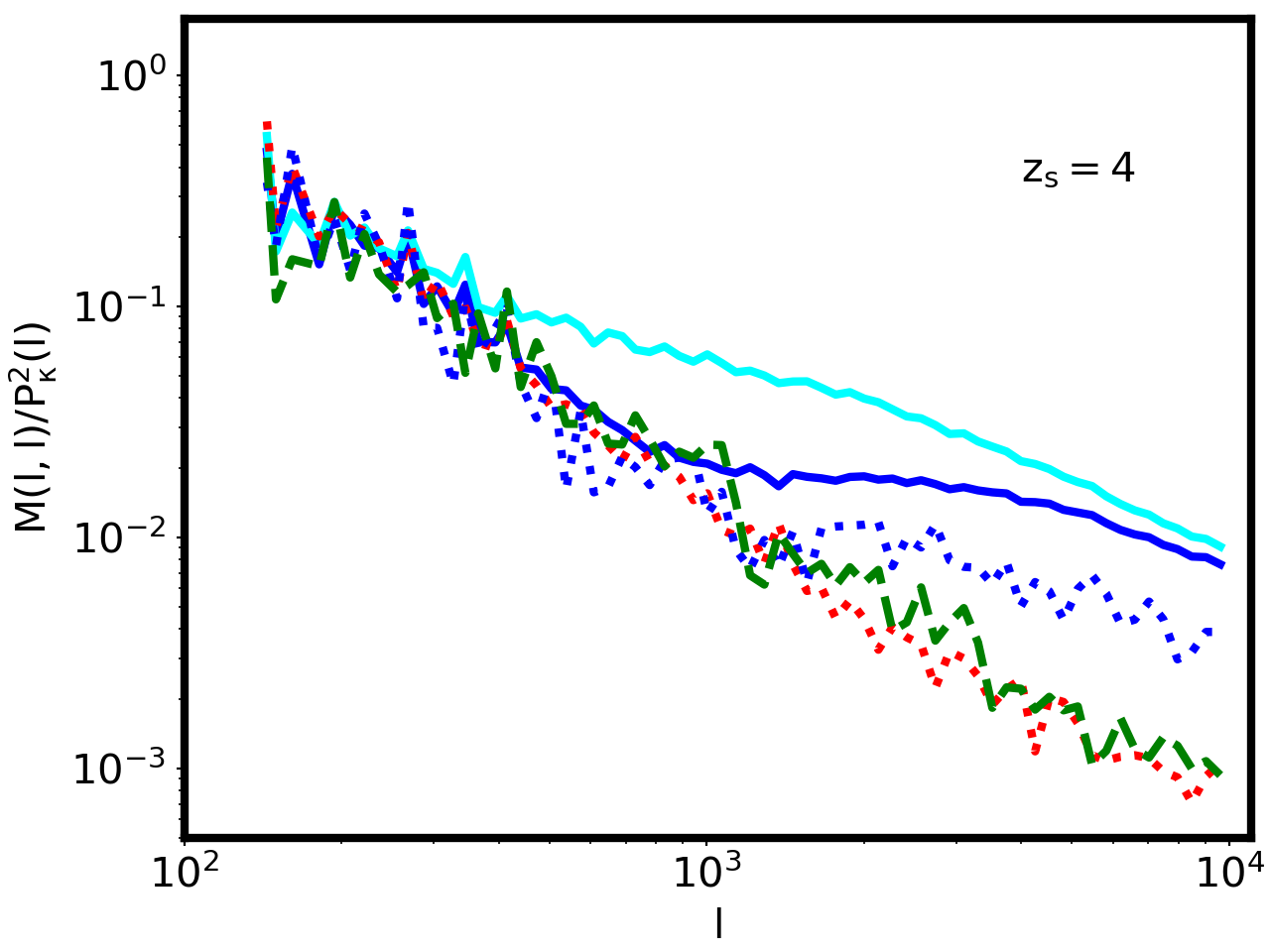}
\caption{Top panels: weak lensing covariance matrices for $z_s=0.5$,
  $1.4$ and $4$ from left to right, respectively, constructed from 512
  past-light-cone \textsc{pinocchio} simulations -- 3LPT runs with a
  grid of $1024^3$ -- and using our fast weak lensing halo model
  \textsc{wl-moka}. The aperture of the field of view is set to 5 deg
  on a side and convergence maps are resolved with $2048^2$
  pixels.\label{covariances} Bottom panels: signal-to-noise of the
  diagonal term of different covariance matrices. In the central panel
  (the closest to the mean of the source redshift distribution useful
  from weak lensing expected from future wide-field-survey) we display
  also the result from Gaussian covariance (dash-dotted orange curve)
  and from the observational discrete number density of background
  sources \citep{refregier04} adopting a Euclid-like sky coverage
  of 15.000 sq. degs. and 30 galaxies per square arcmin (dotted
  magenta).}
\end{figure*}

The inference for cosmological parameters, expressed in term of the
data vector $\vec{\Theta}$, using weak gravitational lensing is found
by minimising the likelihood function that compares observational data
to reference models
\citep{takada04,simon04,kilbinger13,kitching14,kohlinger17}.  In the
case of Gaussian distributed data the likelihood can be approximated
as:
\begin{eqnarray}
  \mathcal{\ln L}(\vec{\Theta}) \propto - \dfrac{1}{2} \sum_{i,j=1}^{N_{b}} \left[
    P^{``obs''}_{\kappa}(l_i) - P_{\kappa}(l_i,\vec{\Theta})
    \right] M(l_i,l_j)^{-1} \nonumber \\
  \left[ P^{``obs''}_{\kappa}(l_j) - P_{\kappa}(l_j,\vec{\Theta}) \right] \,,
\end{eqnarray}
where $M(l,l')$ represents the covariance matrix; its inverse is
usually termed precision matrix. It is worth noting that in the
equation above we have neglected the cosmological dependence of the
covariance matrix. However, this is the subject of ongoing study and
discussion \citep{labatie12,harnois-deraps19} and we plan to address
it in a future work focused both on clustering and weak gravitational
lensing.

\begin{figure*}
\includegraphics[width=0.5\hsize]{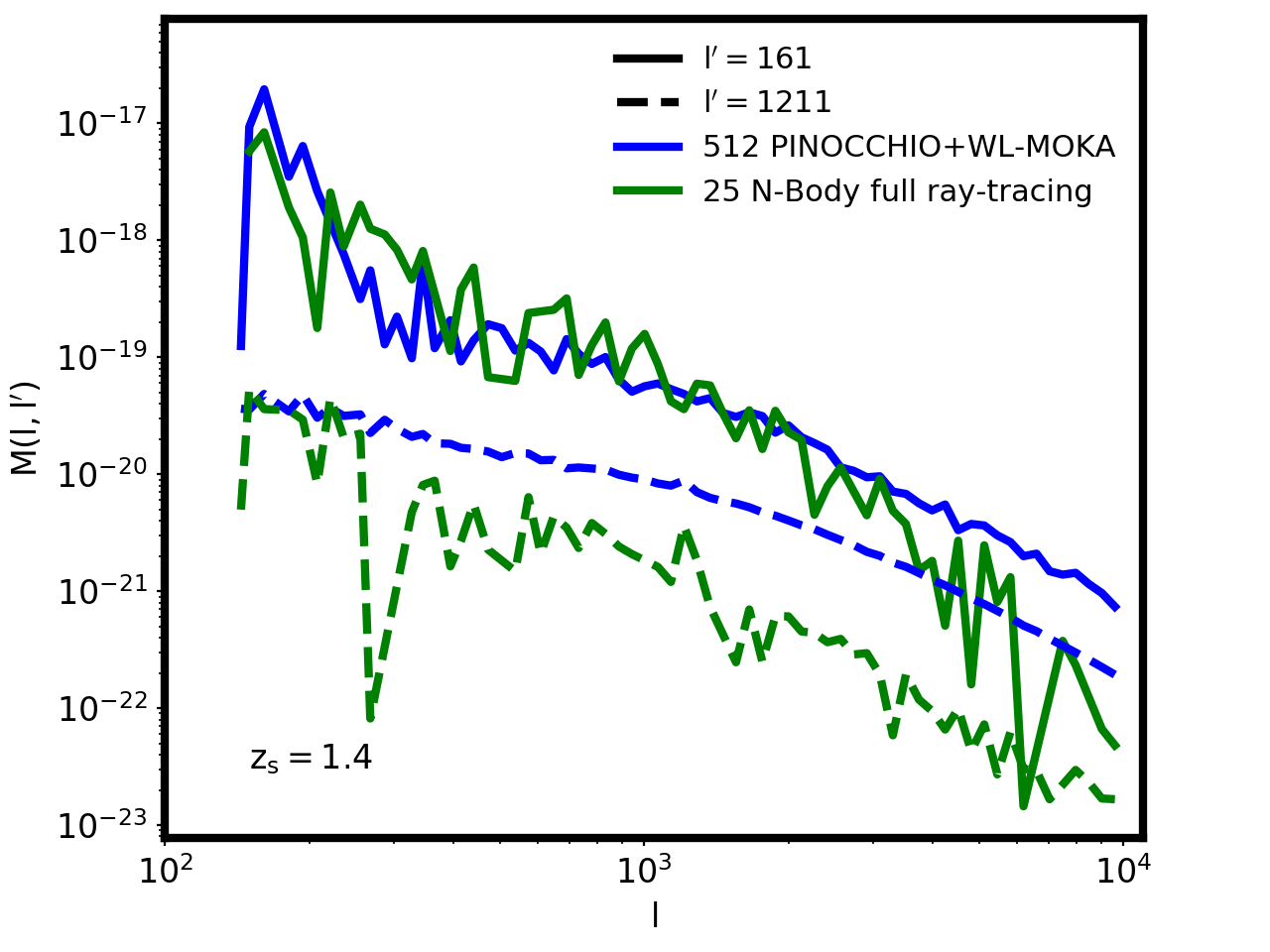}
\includegraphics[width=0.48\hsize]{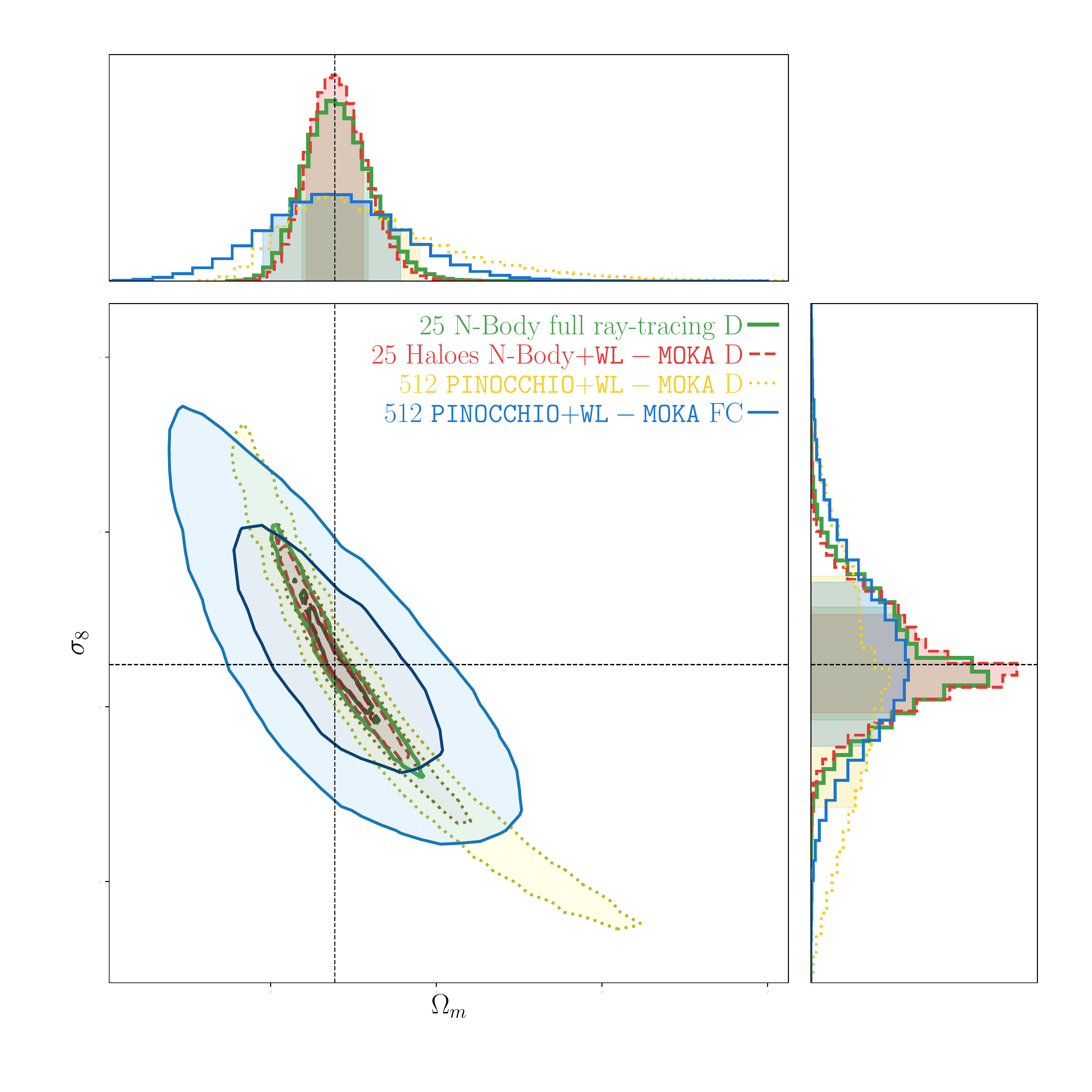}
\caption{Left panel: comparison between off-diagonal covariance matrix
  terms, at two fixed values for $l'$, computed from $25$ N-Body full
  ray-tracing simulations and $512$ simulations obtained using the
  approximate methods for $z_s=1.4$. Right panel: 1 and 2 $\sigma$
  cosmological constraints in the $\Omega_m$-$\sigma_8$ parameter
  space computed adopting different covariances for $z_s=1.4$. Green
  and red contours adopt the diagonal term of the covariances
  constructed from $25$ N-Body convergence map realisations using
  particle full ray-tracing and halo population plus \textsc{WL-MOKA}
  as in \citet{giocoli17}, respectively. The yellow and blue contours
  make use of the covariance from $512$ approximate methods
  convergence maps: considering only the diagonal (D) or the full
  covariance (FC), respectively.  Being interested in the relative
  performances of the various covariance terms, we follow the same
  approach as in \citet{krause17} not including the axis
  values.\label{lastcosmo}}
\end{figure*}

Using different weak lensing light-cone realisations we can build the
covariance matrix as:
\begin{equation}
M(l,l') = \langle (P_{\kappa}(l) - \bar{P}_{\kappa}(l))
(P_{\kappa}(l') - \bar{P}_{\kappa}(l') )\rangle
\end{equation}
where $\langle \bar{P}_{\kappa}(l)\rangle$ represents the best
estimate of the power spectrum at the mode $l$ obtained from the
average, or the median, of all the corresponding light-cone
realisations and $P_{\kappa}(l)$ represents the measurement of one
realisation.  The matrix is then normalised as follows:
\begin{equation}
m(l,l') = \dfrac{M(l,l')}{\sqrt{M(l,l) M(l',l')}}\,,
\end{equation}
in order to be unity on the diagonal.  The covariance matrix
constructed in this way accounts both for a Gaussian and non-Gaussian
contribution arising from mode coupling due to non-linear clustering
and for the survey geometry
\citep{scoccimarro99,cooray01,harnois-deraps12,sato13}.  Off-diagonal
terms with value near unity indicate high correlation while values
approaching zero indicate no correlation.  The covariance matrices
using the convergence power spectra of the maps generated using PLC
halo catalogues from {\sc pinocchio} account for correlation between
observed modes and those with wavelength larger than the simulated
field of view, or survey size.  In fact ($i$) the simulations have $1$
Gpc$/h$ comoving box side, ($ii$) the simulated convergence maps have
been constructed from a cone with aperture $7.2$ deg, ($iii$) each PLC
halo catalogue, produced using {\sc pinocchio}, originated from a
different initial condition realisation.  We underline also that our
covariances do not account for any noise nor systematic errors that
typically enter in the uncertainty budget for the cosmological
forecasts \citep{fu08,hildebrandt17}.

The top panels of Fig. \ref{covariances} we show the covariance matrix
for the convergence power spectra computed from our reference 512,
$1024^3$ mesh grid considering the 3LPT method to displace halo and
particle positions. The three panels refer to the covariances for
sources at $z_s=0.5$, 1.4 and 4 from left to right, respectively. From
the figure we note very good quantitative agreement with the results
that have been presented on the same field of view \citep{giocoli17}
and qualitatively with the covariances computed by \citep{giocoli16a}
on the BigMultiDark \citep{prada16} light-cones that have a
rectangular geometry matching the W1 and W4 VIPERS fields
\citep{guzzo14}. In Appendix \ref{appx1} we show the covariance
matrices, at the same three fixed source redshifts, using only haloes,
in building the different convergence maps, which display appreciable
differences -- with respect to the ones presented here -- mainly for
sources at higher redshifts. In the bottom panels of
Fig. \ref{covariances} we display the signal-to-noise of the diagonal
terms of various covariances, for comparison purpose. The green dashed
curves refer to the realisations obtained from $25$ convergence maps
of our reference N-Body simulation, the solid blue and cyan curves
represent the ones obtain from $512$ runs of {\sc pinocchio} halo
catalogues plus {\sc wl-moka} adding or not the diffuse matter
component among haloes using linear theory predictions,
respectively. In the central panel (the closest source redshift to the
mean of the redshift distribution useful from weak lensing expected
from future wide-field-surveys) we show the Gaussian covariance
(orange dot-dashed) -- obtained running 10.000 Gaussian realisation of
the convergence power spectrum -- and the (dotted magenta)
contribution due to the discrete number density of background sources
\citep{refregier04} adopting a Euclid-like sky coverage of $15.000$
sq. degs. and $30$ galaxies per square arcmin. From the figure we can
notice that the diagonal term of the covariance computed from $512$
{\sc pinocchio} PLC halo catalogues (blue solid curve) is at large
scale in agreement with the full try-tracing simulations while at
small scales it moves up dominated by sample variances
\citep{barreira18a,barreira18b}. All $512$ runs have different initial
conditions, while the $25$ full ray-tracing convergence maps are
extracted from the same simulation with fixed IC. At large angular
modes the blue curves move down because of a smaller number of PLC
realisations.

A systematic comparison of the off-diagonal term contributions of the
covariance, for $z_s=1.4$, is displayed in Fig.~\ref{lastcosmo}. In
the left panel we show the terms at two fixed values for $l'$ of the
covariances computed (blue curves) using $512$ realisation maps
adopting our approximate methods and (green curves) employing $25$
full ray-tracing maps from our reference N-Body simulation. From the
figure, we notice that for intermediate values for $l'$ the two
covariances are relatively comparable, while for larger values of $l$
they deviate mainly due to both numerical resolution and particle
noise contributions. As underlined in \citet{giocoli17}, the full-ray
tracing covariances are quite noisy and problematic to be inverted
when used to constrain cosmological parameters. In the right panel we
show the cosmological constraints obtained adopting various
covariances in the $\Omega_m$-$\sigma_8$ parameter space. The results
refer again to $z_s=1.4$. Since the aim of this test is to display the
relative performances of the various covariance terms, we follow the
same approach as in \citet{krause17} not including the axis values. A
self consistent analysis to constraint cosmological parameters from
weak lensing datasets could not neglect shape measurements inaccuracy
\citep{berstein02,kacprzak12,miller13}, photometric redshift errors
and uncertainties of the reconstructed source redshift distribution
\citep{hildebrandt12,benjamin13,yao19,wright20}.  The green and red
contours show the cosmological constraints derived when adopting the
covariance matrices constructed from $25$ realisations of the
projected density field up to $z_s=4$ using full ray-tracing particle
simulations and the halo catalogues plus \textsc{WL-MOKA} as in
\citet{giocoli17}, respectively. Since the covariance matrix from $25$
full ray-tracing simulations is noisy we could use only the diagonal
(D) term. We underline that the Gaussian covariance gives similar
constraints to the green contours, not shown to avoid overcrowding the
figure. The yellow contours show the constraints derived when using
the diagonal term of the covariance constructed from $512$ approximate
methods simulations, while the blue ones adopt the full covariance
matrix, which exhibit a degradation of the Figure of Merit.  The
cosmological constraints have been obtained implementing the modelling
of the cosmic shear power spectrum in the CosmoBolognaLib
\citep{marulli15} and accounting for the number of realisations when
constructing the covariance as described by \citet{hartlap07} and
\citet{percival14}. We remind the reader that the use of these
approximate covariances for analysing existing and future wide field
surveys will require more consistent tests with theoretical models,
path we are currently pursuing inside different collaborations.

\section{Summary \& Conclusions}
\label{summary}
In this paper we have presented a natural extension of the approximate
Lagrangian perturbation theory code {\sc pinocchio} dedicated to
creating fast and accurate convergence maps for weak gravitational
lensing simulations. Since the methods implemented are quite general
they constitute a tool for full cosmological analysis of observational
data-sets going from galaxy clustering to cluster counts and
clustering to cosmic shear.

The main points of this work are:

- the halo mass function in {\sc pinocchio} past-light-cones is in
very good agreement with both numerical simulation data and
theoretical models with which we compare to;

- the expected convergence power spectra constructed from our
reference runs using only haloes, present within the past light-cone,
are quite well recovered on small scales, however we need to include
also the contribution from matter present outside haloes to fully
reconstruct the large scale modes as predicted from linear theory,
this has been discussed and motivated with a dedicated comparison with
the analytical halo model for non-linear power spectrum;

- the full convergence maps have a power spectrum that is in agreement
well within $5\%$ of that obtained from full ray-tracing through
light-cones constructed from the reference cosmological N-Body
simulation;

- the contribution of galaxy clusters to the total convergence power
spectra at different source redshifts for $l<3\times 10^3$ remains
constant, deviating approximately by $30\%$ with respect to the total
ones, with a slight evolution with redshift due to the rarity of
clusters at large look back times;

- the weak lensing power spectra obtained running {\sc pinocchio} with
2LPT displacement and then {\sc wl-moka} agree within $1\%$ with the
3LPT reference run, however when using ZA the projected power spectra
deviate more than $7\%$ on large angular modes;

- the relative differences between the convergence power spectra of
our fast methods for weak lensing with respect to the reference
measurements from N-body simulations is well below $5\%$, consistent
with what has been found also by other approximate methods;

- the speed of our algorithms allows for the possibility of generating a
very large number of light-cone weak lensing simulations and the
opportunity to construct self-consist covariances for weak
gravitational lensing. 

A fast and accurate method for generating convergence maps using
approximate methods is needed in light of the expected data from
future wide field surveys. In this work we have presented the
interfacing of {\sc pinocchio} and {\sc wl-moka} which enables them to
simulate cosmic shear signals from large scale structures.  This adds
a new capacity to {\sc pinocchio}, beyond the halo mass function and
clustering, applicable to simulate quickly and consistently various
covariances for different statistics, opening a new window for robust
cosmological analyses of future observational data-sets.

\section*{Acknowledgments}
CG and MB acknowledge support from the Italian Ministry for Education,
University and Research (MIUR) through the SIR individual grant
SIMCODE, project number RBSI14P4IH.  We acknowledge the grants ASI
n.I/023/12/0, ASI-INAF n.  2018-23-HH.0 and PRIN MIUR 2015 Cosmology
and Fundamental Physics: illuminating the Dark Universe with Euclid".
CG, LM and MM are also supported by PRIN-MIUR 2017 WSCC32 ``Zooming
into dark matter and proto-galaxies with massive lensing
clusters''. TC is supported by the INFN INDARK PD51 grant.  We
acknowledge the anonymous reviewer for his/her useful comments that
help improving the presentation of our methods and results. CG is
grateful to Alex Barreira, Sofia Contarini, Wolfgang Enzi, Federico
Marulli and Alfonso Veropalumbo for helpful discussions and comments.

\appendix

\section{Halo Covariance Weak Lensing Power Spectra}
\label{appx1}
\begin{figure*}
\includegraphics[width=0.33\hsize]{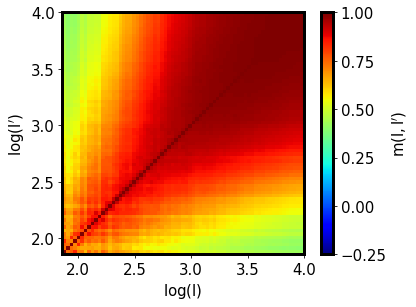}
\includegraphics[width=0.33\hsize]{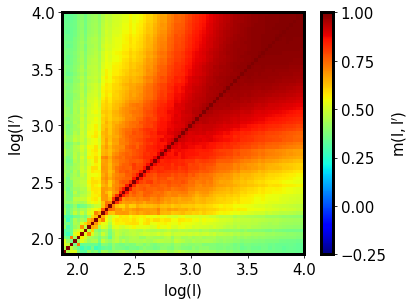}
\includegraphics[width=0.33\hsize]{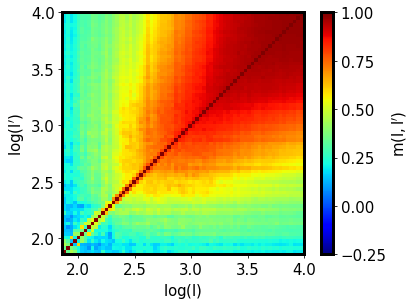}
\caption{Same as Fig.~\ref{covariances} but considering in our
  convergence maps only the halo contributions: {\sc pinocchio} plus
  {\sc wl-moka} (only haloes). The three panels display the
  covariance matrix for sources at $z_s=0.5$, $1.4$ and $4$ from
  left to right, respectively. \label{covarianceshaloes}}
\end{figure*}

In Figure~\ref{covarianceshaloes} we display the covariance matrices
of the convergence power spectra from $512$ different maps build using
only haloes, i.e. on large scale the convergence power spectra are not
forced to follow linear theory. The three panels from left to right
show the convergence power spectrum covariance at three different
source redshifts: $z_s=0.5$, $1.4$ and $4$.

\section{Convergence Maps at Different Grid Resolutions}
\label{appx2}
\begin{figure*}
    \includegraphics[width=0.75\hsize]{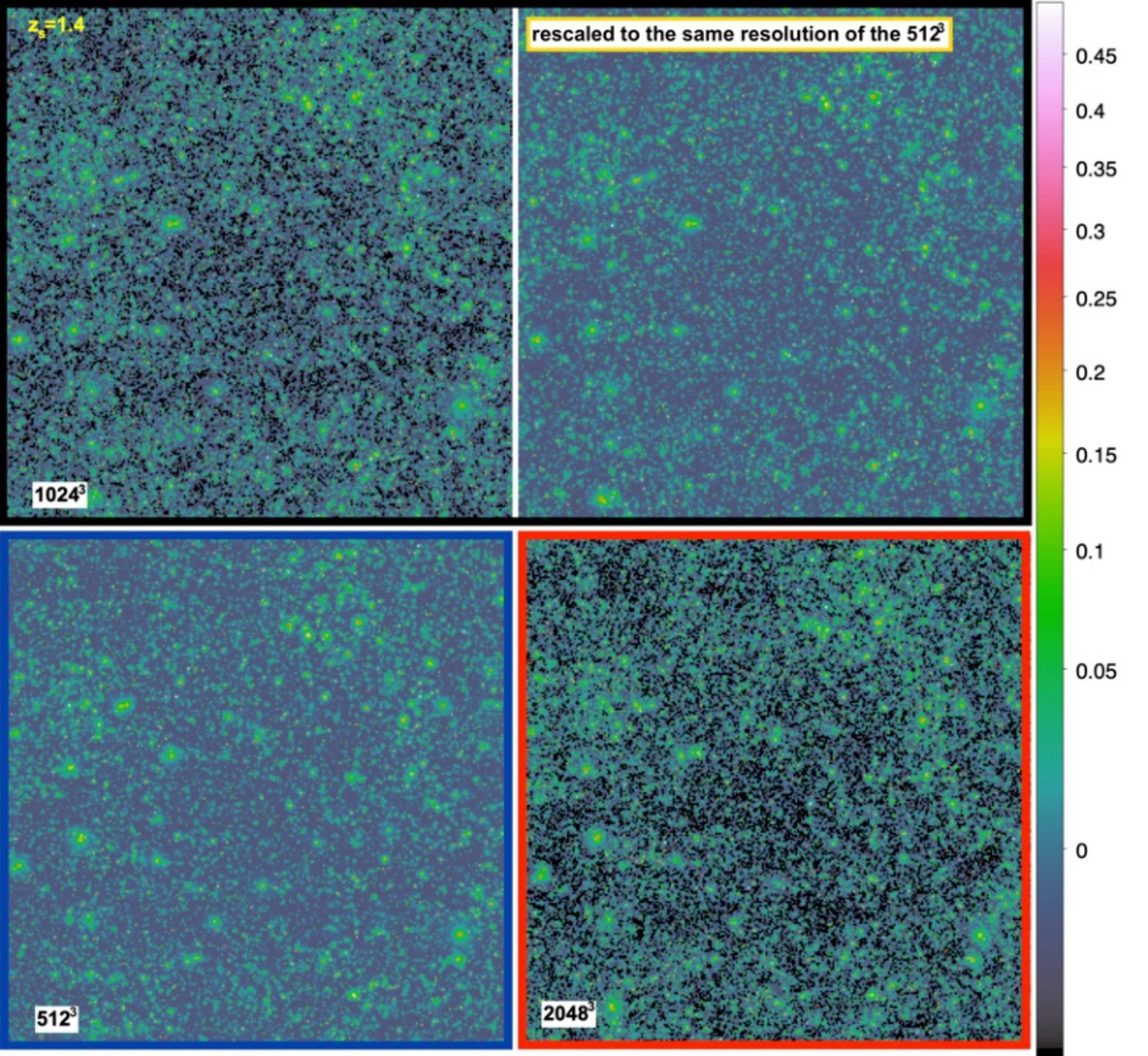}
    \caption{Convergence maps at $z_s=1.4$ for the same randomisation
      but for different runs. Top left, bottom left and bottom right
      display the convergence maps for the {\sc pinocchio} runs with
      $1024^3$, $512^3$ and $2048^3$ particles, top right shows the
      convergence map build with running {\sc wl-moka} on the $1024^3$
      run but considering all haloes with mass larger than $5.6\times
      10^{12}M_{\odot}/h$, which corresponds to the halo mass
      resolution of the $512^3$.  In all cases we have adopted 3LPT
      displacement field and all maps are resolved with the same
      resolution of $2048\times 2048$ pixels on an angular scale of 5
      deg by side.
    \label{figkappares}}
\end{figure*}

In this section we display the convergence maps constructed adopting
different particle and grid resolutions when running {\sc pinocchio}
while using the same seed and phases when displacing particles and
haloes (3LPT) from the initial conditions.  Fig. \ref{figkappares}
shows the convergence map for sources at $z_s=1.4$; top left, bottom
left and bottom right panels exhibit the results when running {\sc
  wl-moka} on the PLC halo catalogues of the $1024^3$, $512^3$ and
$2048^3$ {\sc pinocchio} simulations, respectively.  In all cases, we
notice, the large scale structure is similar but the haloes have
different displacements. The top right panel shows the convergence map
constructed on the halo catalogue of the $1024^3$ run, but considering
only systems more massive than $5.6\times 10^{12}M_{\odot}/h$, which
corresponds to the halo mass resolution of the $512^3$. We recall for
the reader that when building the convergence maps we assume the
average to be zero \citep{hilbert19} and that they are resolved with
$2048 \times 2048$ pixels on an angular scale of 5 deg on a side.

\begin{figure}
    \includegraphics[width=0.9\hsize]{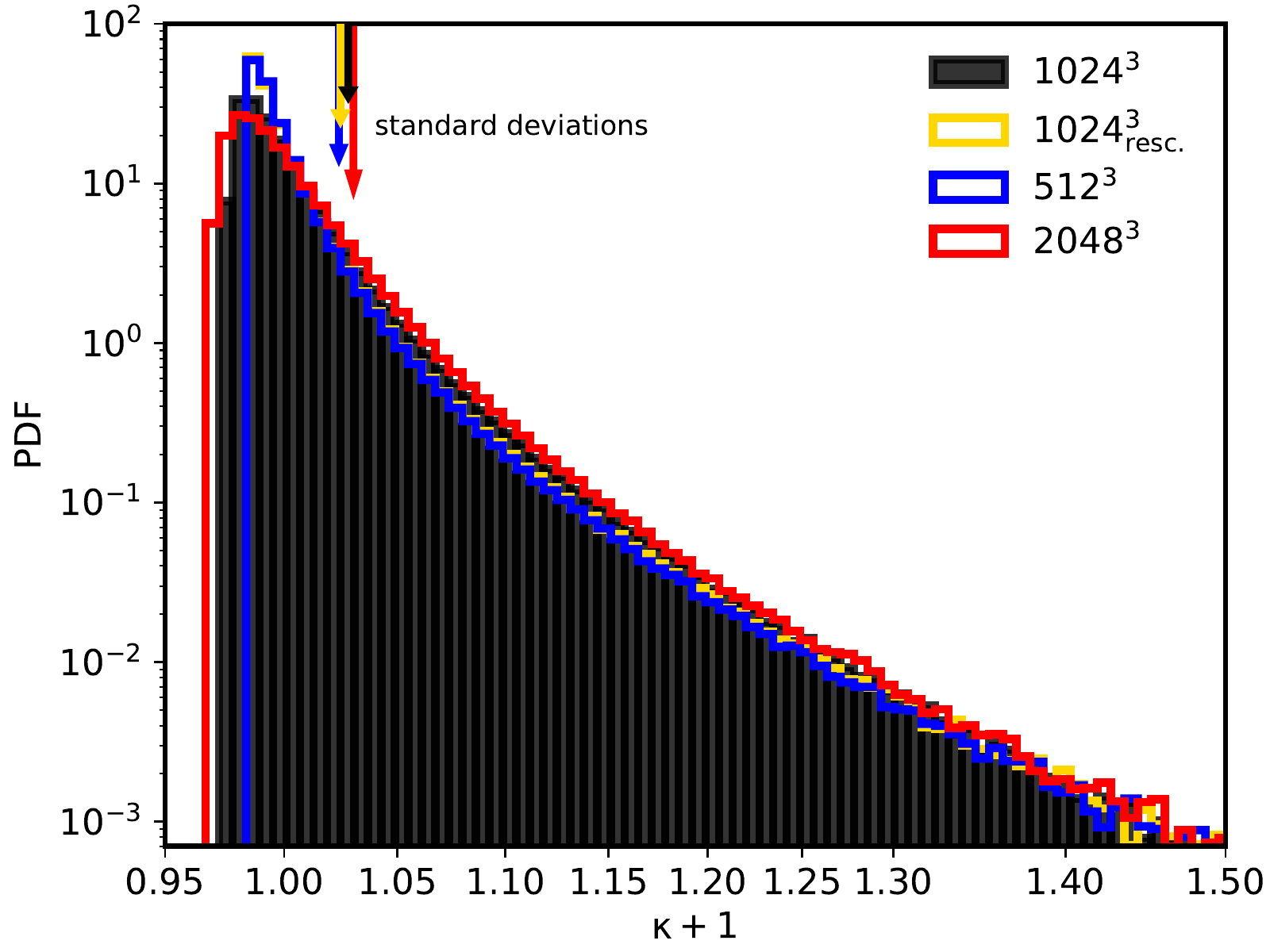}
    \caption{Probability Distribution Function per pixel of the
      convergence maps presented in Fig.~\ref{figkappares}.  The
      arrows mark the standard deviation of the distributions, we
      recall the reader that by construction the maps have average
      equal to zero \citep{hilbert19}.
    \label{figkapparespdf}}
\end{figure}

In Fig. \ref{figkapparespdf} we show the normalised Probability
Distribution Function per pixel in the maps presented in
Fig. \ref{figkappares}. The different colours refer to the various
runs and the arrows indicate the standard deviations of the
distributions.

\bibliographystyle{mn2e}

\bsp
\bibliography{globalbibs.bib}
\label{lastpage}
\end{document}